\documentclass[12pt]{iopart}
\usepackage{graphics}
\usepackage{flushend}
\usepackage{cuted}
\usepackage{subfigure}
\usepackage[dvips]{graphicx}
\usepackage{color}
\usepackage{sidecap}

\newcommand{\be}{\begin{equation}}
\newcommand{\ee}{\end{equation}}

\newcommand{\bea}{\begin{eqnarray}}
\newcommand{\eea}{\end{eqnarray}}
\newcommand{\bd}{\begin{displaymath}}
\newcommand{\ed}{\end{displaymath}}
\newcommand{\ba}{\begin{array}}
\newcommand{\ea}{\end{array}}
\newcommand{\ei}{\end{itemize}}
\newcommand{\bc}{\begin{center}}
\newcommand{\ec}{\end{center}}
\newcommand{\bfl}{\begin{flushleft}}
\newcommand{\efl}{\end{flushleft}}
\newcommand{\bfr}{\begin{flushright}}
\newcommand{\efr}{\end{flushright}}

\def\bR{{\bf R}}
 \def\bq{{\bf q}}

\def\6{\partial}

  \def\S{\Sigma}

\def\={\!\!\!&=&\!\!\!}
\def\+{\!\!\!&&\!\!\!+~}
\def\-{\!\!\!&&\!\!\!-~}

\begin{document}

\title[Evolution of  the multiband RKKY interaction]{Evolution of  the multiband RKKY interaction:   Application to   iron pnictides and chalcogenides}

\author{Alireza Akbari$^{1}$, Peter Thalmeier$^{1}$, and Ilya Eremin$^{2}$}

\address{$^{1}$ Max-Planck Institute for the  Chemical Physics of Solids, D-01187 Dresden, Germany \\
   $^{2}$ Theoretische Physik III, Ruhr-Universit\"{a}t Bochum, D-44780, Bochum, Germany 
}

\ead{akbari@cpfs.mpg.de}
\begin{abstract}
The indirect RKKY interaction in  iron pnictide and chalcogenide metals  is calculated for
a simplified four bands Fermi surface (FS) model. We investigate the specific multi-band features  
and show that distinct length scales of the RKKY oscillations appear.
For the regular lattice of the local moments, the generalized RKKY interaction is defined in momentum space.
We consider its momentum dependence in paramagnetic  and spin density wave (SDW) phases, discuss its implications for the possible type of magnetic order and compare
it to the results obtained from more realistic tight-binding type Fermi surface model.
Our finding can give important clues on  the magnetic ordering of the 4f- iron  based superconductors. 
\end{abstract}
\pacs{74.70.Xa, 75.30.Hx,75.30.Fv}
\maketitle
\section{Introduction}
\label{sect:introduction}

The discovery of  iron-based superconductors  \cite{Kamihara:2008fk}, has lead to a
renewed interest in multiband superconductivity  \cite{Ishida:2009,Stewart:2011}.
Their parent compounds are paramagnetic metals at high temperature and  mostly show an antiferromagnetic (AF) order of spin density wave (SDW)  type slightly below a tetragonal to orthorhombic structural transition  \cite{Lumsden:2010}. Magnetic order of the itinerant Fe moments is driven by the nesting properties of 3d type Fermi surfaces which consist of hole and electron pockets around the center and at the boundaries of the Brillouin zone (BZ). The magnetism of the parent  compounds becomes even more involved in rare earth (R) based Fe-pnictides where layers with localized 4f moments exist that are separated from the Fe-As layers with itinerant 3d moments.  It is found that the R- moments order at temperatures much below the SDW transition. Their magnetic structure may be generally different from that of the Fe layers.  The latter predominantely show the collinear in-plane stripe structure of Fe moments while the R layers may order  in different noncollinear or simple ferromagnetic structure.
For example the in-plane FM order appears in the 122 family RFe$_2$As$_2$ for R = Eu\cite{Xiao:09}, and in the 1111 family RFeAsO the R = Ce compound shows  non-collinear in-plane stripe structure\cite{Zhao:2008b,Maeter:2009}.

 Because of the close connection of magnetic and electronic properties in these 
kind of  materials, investigation of 
magnetic order  and effective coupling mechanism of local moments   is a question of great interest.
The latter may belong to a periodic sublattice of 4f-elements \cite{Pourovskii:2008,Jesche:2009,Maeter:2009,Xiao:09};
doped disordered impurity magnetic moments  (diluted magnetic moments) \cite{Kitchen:2006,Texier:2012,Alfonsov:2012},
or possibly partially localized 3d electrons on the Fe sublattice originating from an orbitally selective Mott-Hubbard localization.

A well studied example of the 4f- based 122 Fe pnictides  is  EuFe$_{2}$As$_{2}$, with its highest SDW transition temperature  reported
 at $T_{SDW}=190$K   \cite{Xiao:09,Jeevan:2008,Ren:2009,Jiang:2009,Jeevan:2011,Tokiwa:2012}.
The AF ordering   of localized  4f- moments (Eu$^{2+}$ with $S=7/2$) happens at lower temperature, $T_N\approx 20\mbox{K}$, for a {\it different} wave vector ${\bf Q}=(0,0,1)$.
This means the Eu$^{2+}$-moments  are FM ordered in  $ab$-planes which is in contrast to the columnar AF order of Fe itinerant moments \cite{Xiao:09}. In addition,  the relaxational behavior of Eu$^{2+}$ spins in ESR  shows a distinct magnetic anisotropy  below $T_{SDW}$ \cite{Dengler:10}.
The magnetization anisotropy of Eu spins has a temperature dependency which is changing across the SDW transition temperature \cite{Zapf:11}. These observations indicate that the SDW transition in the Fe-3d itinerant subsystem may have important effects on the
effective coupling of local 4f moments. 
Evidence for the coupling of the $4f^7$-electron and conduction electrons is  also obtained by  ESR  in the Gd-based 1111 compounds La$_x$Gd$_{1-x}$FeAsO  (Gd$^{3+}$ with spin S = 7/2) \cite{Alfonsov:2012}.
Below $T\sim 6 $K, neutron scattering measurements also found the Fe-Nd interaction in NdFeAsO single crystals which force the  Nd- moments  to be  ordered AF with the same $ab$-plane configuration as Fe moments\cite{Tian:2010}.

Since the 4f moments are localized, their direct exchange can be  neglected. However one has to consider the indirect  Ruderman-Kittel-Kasuya-Yosida (RKKY)
interaction~\cite{Ruderman:1954vn,Kasuya:1956fr,Yosida:1957zr}  of 4f moments via spin polarization of 3d conduction electrons. The strength of this oscillatory effective exchange is controlled by 
 the distance between two localized 4f moments and the Fermi surface (FS) properties  of 3d conduction electrons.
Such an interaction generally plays an important role in revealing the nature of the magnetism  in metals with partially filled $d$- and $f$-electron shells.
It is well-known that the RKKY interaction for an anisotropic Fermi surface (FS) with nesting consists of several terms originating from flat regions in the conduction bands (van Hove regions) and those which describe the interference
between contributions from their vicinities \cite{Aristov:1997qy}.
For the nearly nested conduction bands the latter term is present down to the
interatomic distances and favors the commensurate antiferromagnetic ordering of the localized moments.
The degree of nesting varies  continuously for different compounds \cite{Yaresko:2009}, where we have a perfect nesting in some pnictides \cite{Coldea:2008} and weakly nested compound of some chalcogenides\cite{Singh:2012,Sadovskii:2012}. 
\\

\begin{figure}
 \centering
\includegraphics[width=0.3 \linewidth]{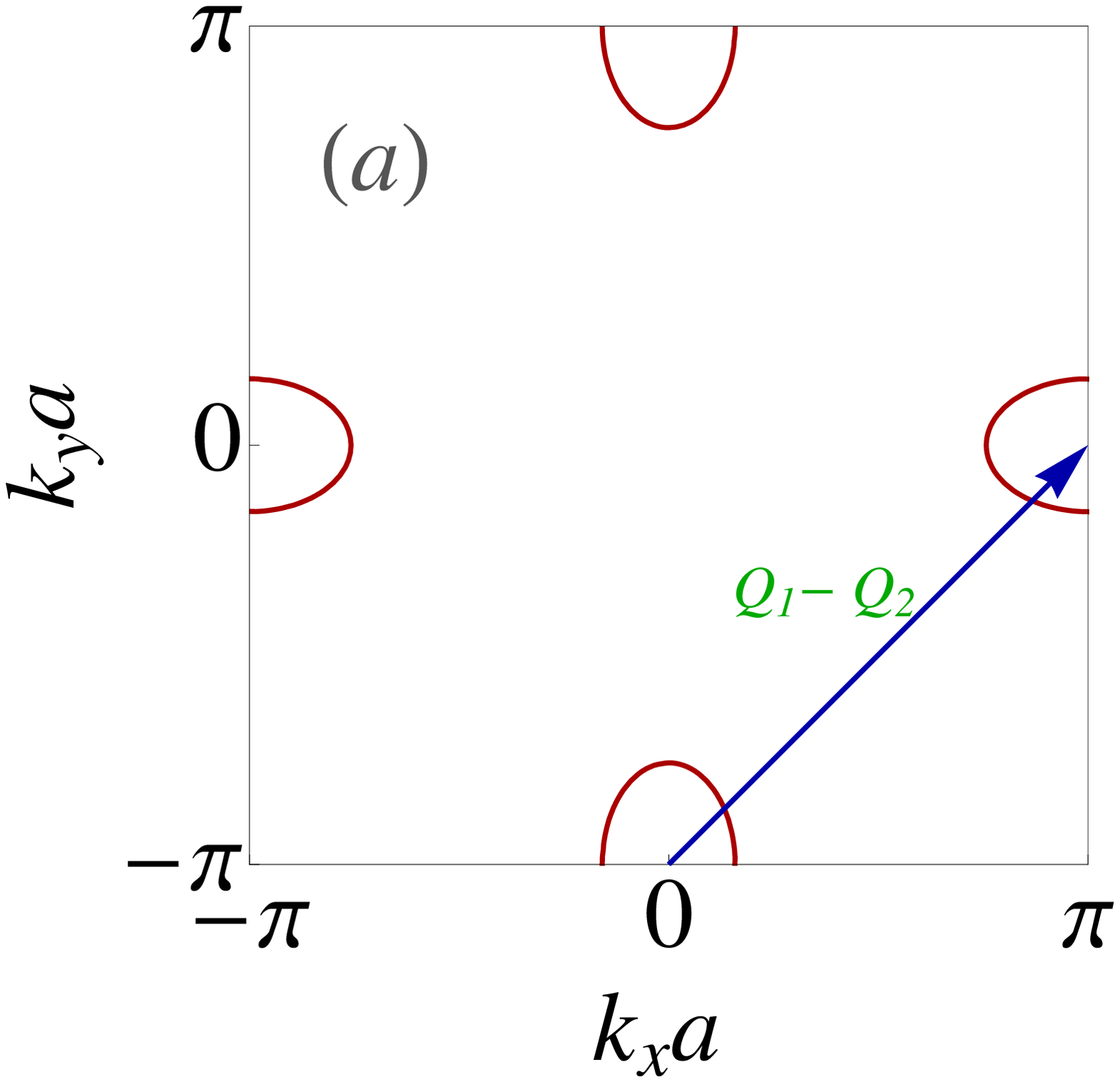}
\includegraphics[width=0.3 \linewidth]{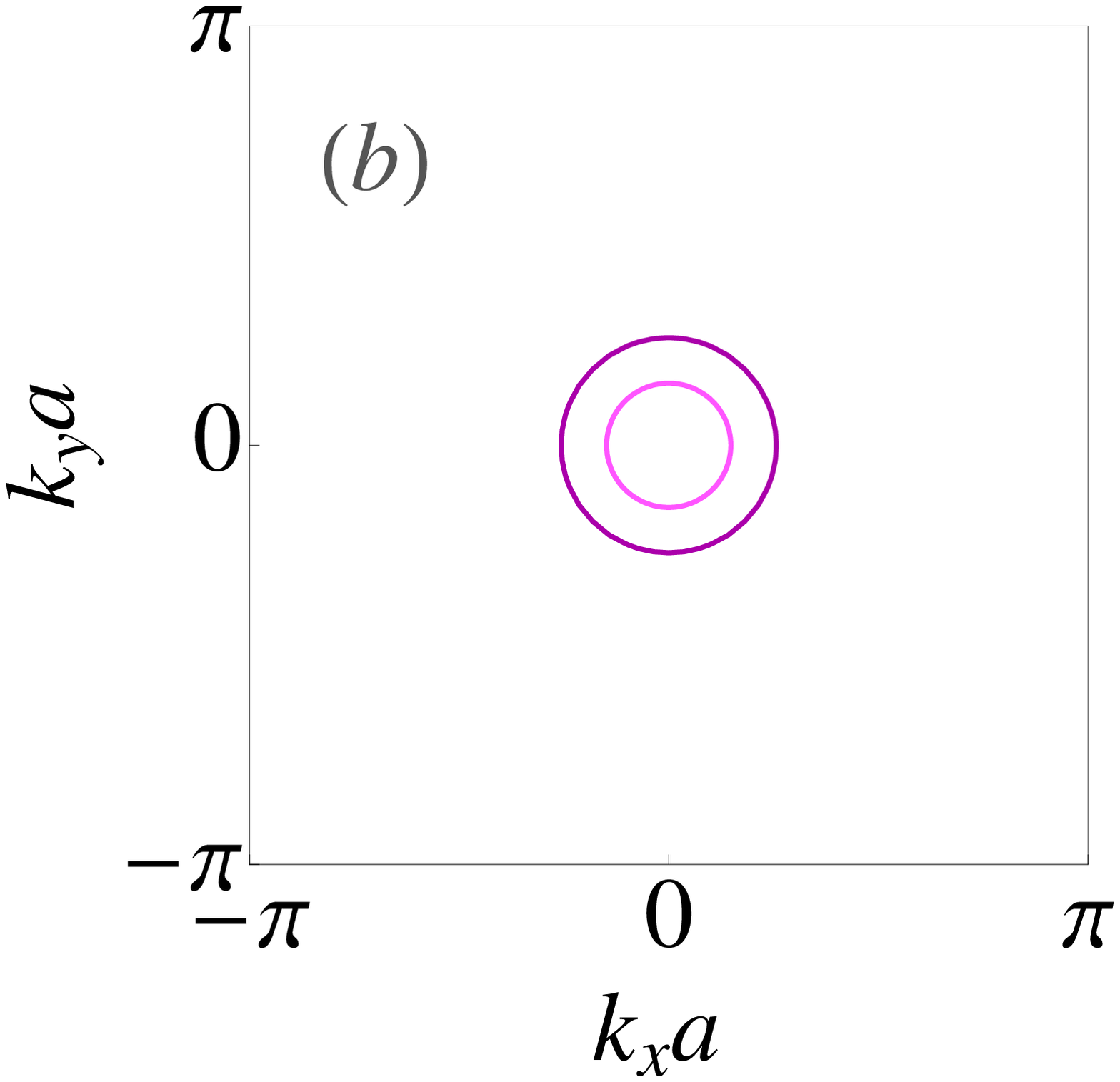}
\includegraphics[width=0.3 \linewidth]{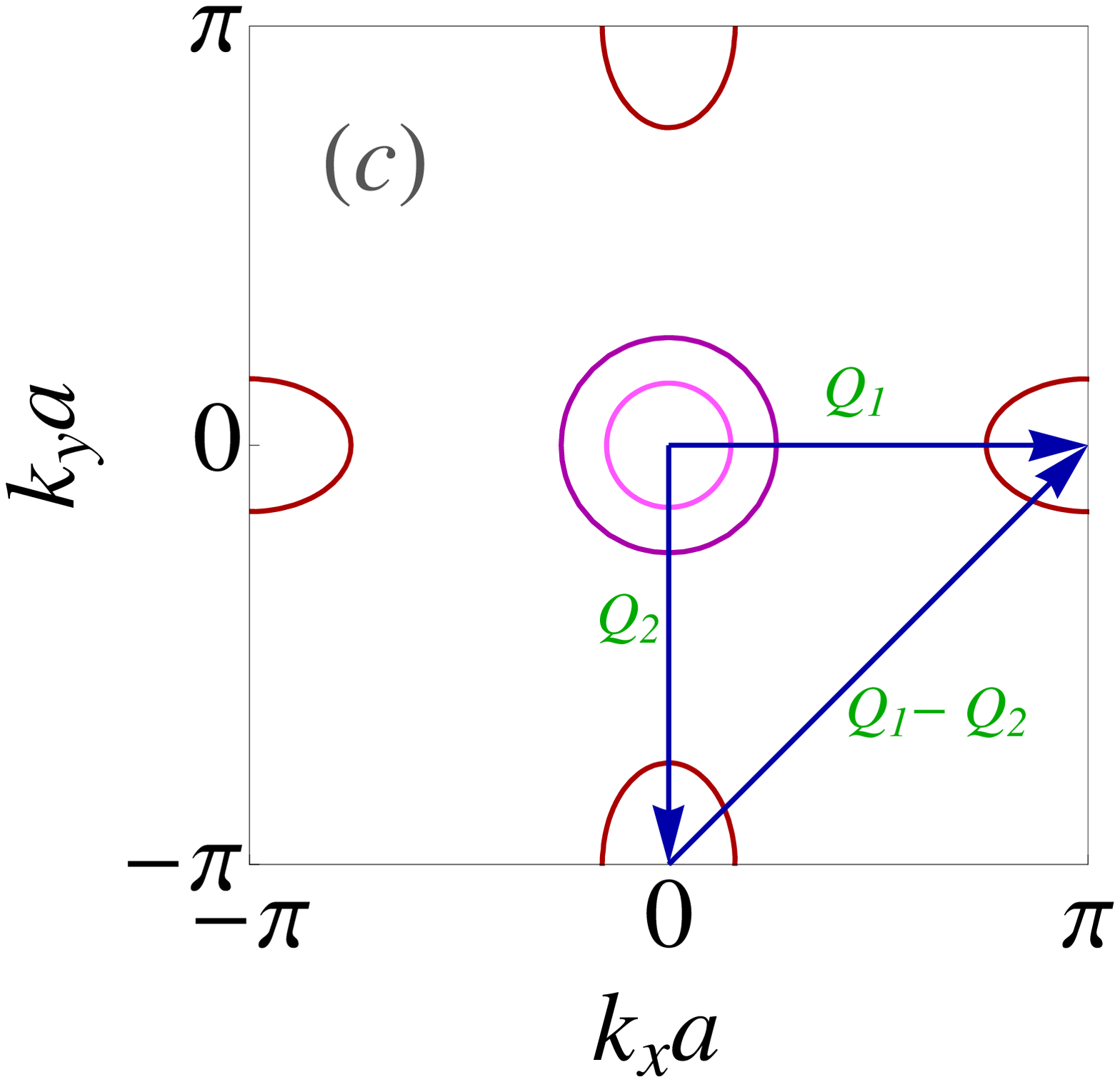}
\caption{(Color online)
a) The Fermi surface with  pure electron- like  pockets with ellipticity $\epsilon=0.4$, centered 
 around the $\mathbf{Q}_\beta$ [$\mathbf{Q}_1=(\pm \pi, 0)$ and $\mathbf{Q}_2=(0, \pm \pi)$] points in the unfolded BZ ($\beta$-bands). b) Fermi surface with pure hole- like  pockets centered at $\Gamma$- point ($\alpha$-bands).
c) Fermi surface with  two circular  hole pockets and two  electron- like  pockets.
Note that  c)  results from  a combination of electron and hole pockets in a) and b).
}
\label{fig1}
\end{figure}
%

This motivates us to investigate how the RKKY mechanism in iron based superconductors depends 
 on the absence or presence of electron- or hole- Fermi surface sheets and their nesting properties. 
 In most of the iron based superconducting  systems, the Fermi surface consists of the hole and electron pockets simultaneously \cite{Lebegue:2007,Singh:2008,Boeri:2008,Mazin:2008,Liu:2008,Evtushinsky:2009,Coldea:2008}.
It is also possible to have a situation with only the hole- like or the electron- like pockets separately.
For example, according to band structure calculations \cite{Xu:2008kx} and angle resolved photoemission spectroscopy (ARPES) \cite{Sato:2009yq}  the overdoped KFe$_2$As$_2$ compound has no electron - like FS sheets \cite{Guo:2010uq}. On the other hand
ARPES studies on the alkali-intercalated Fe chalcogenide system KFe$_2$Se$_2$ have recently shown that there are no hole- like FS sheets for some doping range \cite{Qian:2011qy}.
We will
use a simplified parabolic band model with isotropic hole but anisotropic electron mass and study 
how the  real space oscillations of RKKY local moment pair interaction evolve with such FS change. 
We study the general case involving four bands that consist of  two electron- like and two hole- like sheets.
We compare these results with the special case where only one type of sheet is present and the closed analytical
solution is obtained.

For a lattice of local moments with effective RKKY interactions the latter will also be analyzed in momentum space. This is
relevant for the R-based pnictides with magnetic order appearing both in the itinerant 3d and localized 4f subsystems. 
These results will be compared to a more detailed calculation  using tight-binding type conduction electron bands. 
As indicated by the above examples the SDW transition and associated gap opening may influence the \bq - dependence and anisotropy of the effective RKKY exchange interaction which determines the type of order in the local moment subsystem. This feedback effect on local moment magnetism will also be investigated for the tight binding model.

%
\begin{figure}
 \centering
\includegraphics[width=0.85 \textwidth]{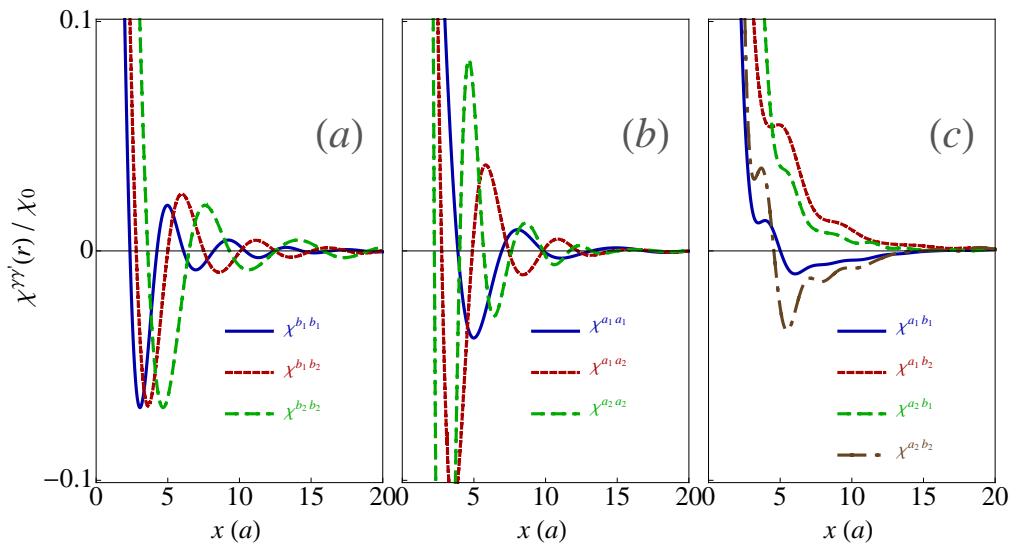}

\caption{(Color online)
The fully analytical calculations of the  individual components  of  magnetic spin susceptibility of conduction electrons  $\chi^{\gamma\gamma^\prime} ({\bf r})$ 
  (obtained using  Eqs.(\ref{chi_gamma_gamma}, \ref{chi_gamma_gammaP}) along the x-direction for a)   pure electron- like pockets,
 b) pure  hole- like pockets.
 They are completely in agreement with the numerical results.
c) The numerical calculation of the individual components of  magnetic spin susceptibility of conduction electrons   $\chi^{\gamma\gamma^\prime} ({\bf r})$  along the x-direction 
 for only the inter hole- and electron- like contributions. 
Note that $\chi^{\gamma \gamma^\prime} ({\bf r}) =\chi^{\gamma^\prime \gamma}  ({\bf r})$,
 furthermore $\chi_0\equiv\frac{m_ea^2}{4\hbar^2}$.
}
\label{fig2}
\end{figure}
%

%
\begin{figure}
 \centering
\includegraphics[width=0.859 \textwidth]{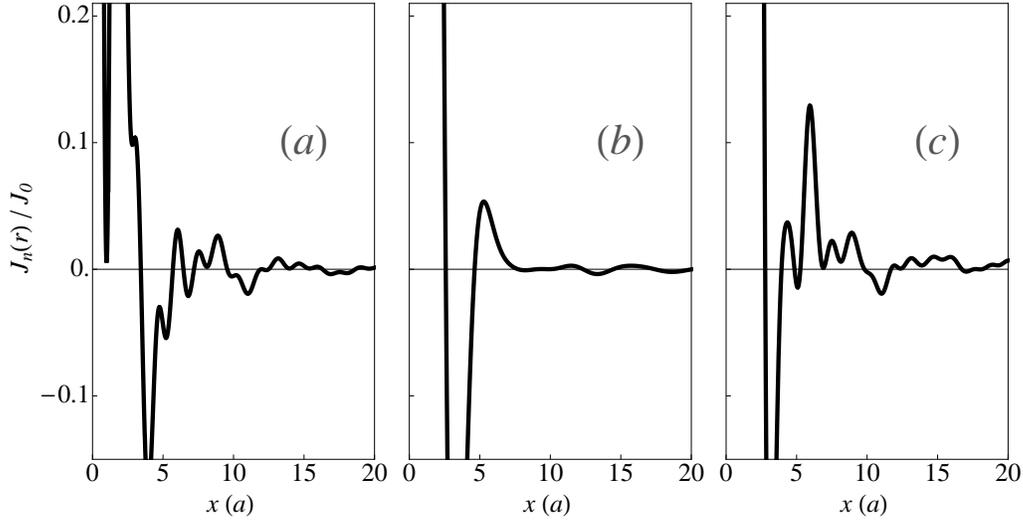}
\caption{
 The normalized total RKKY  interaction,  $J_n({\bf r})/J_0$, for a) pure  electron- like pockets,  b) pure hole- like pockets and c) for the case with 4 electron and hole bands, corresponding to the case of Fig.(\ref{fig1}.c),
  along the x-direction,
 furthermore $J_0\equiv J_{ex}^2\chi_0 $. (a) and (c) exhibit rapid oscillations due to the inter-pocket contributions (c. f. Fig.~\ref{fig1}a,c)
}
\label{fig3}
\end{figure}
%

%
\begin{figure}
 \centering
         \subfigure[]
    {
        \includegraphics[width= 0.27\linewidth]{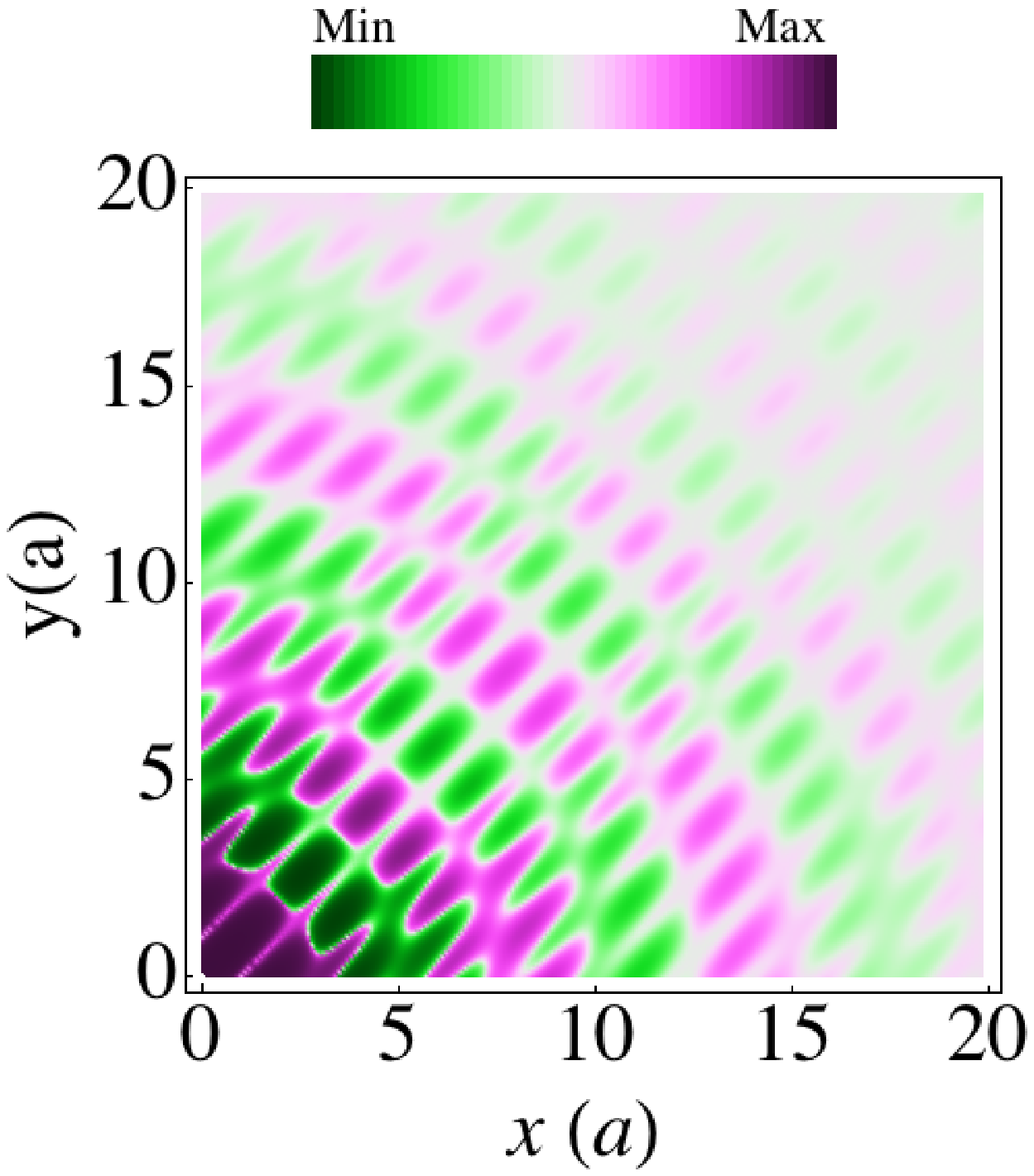}
    }
 \hspace{0.1cm}
          \subfigure[]
    {
        \includegraphics[width= 0.27\linewidth]{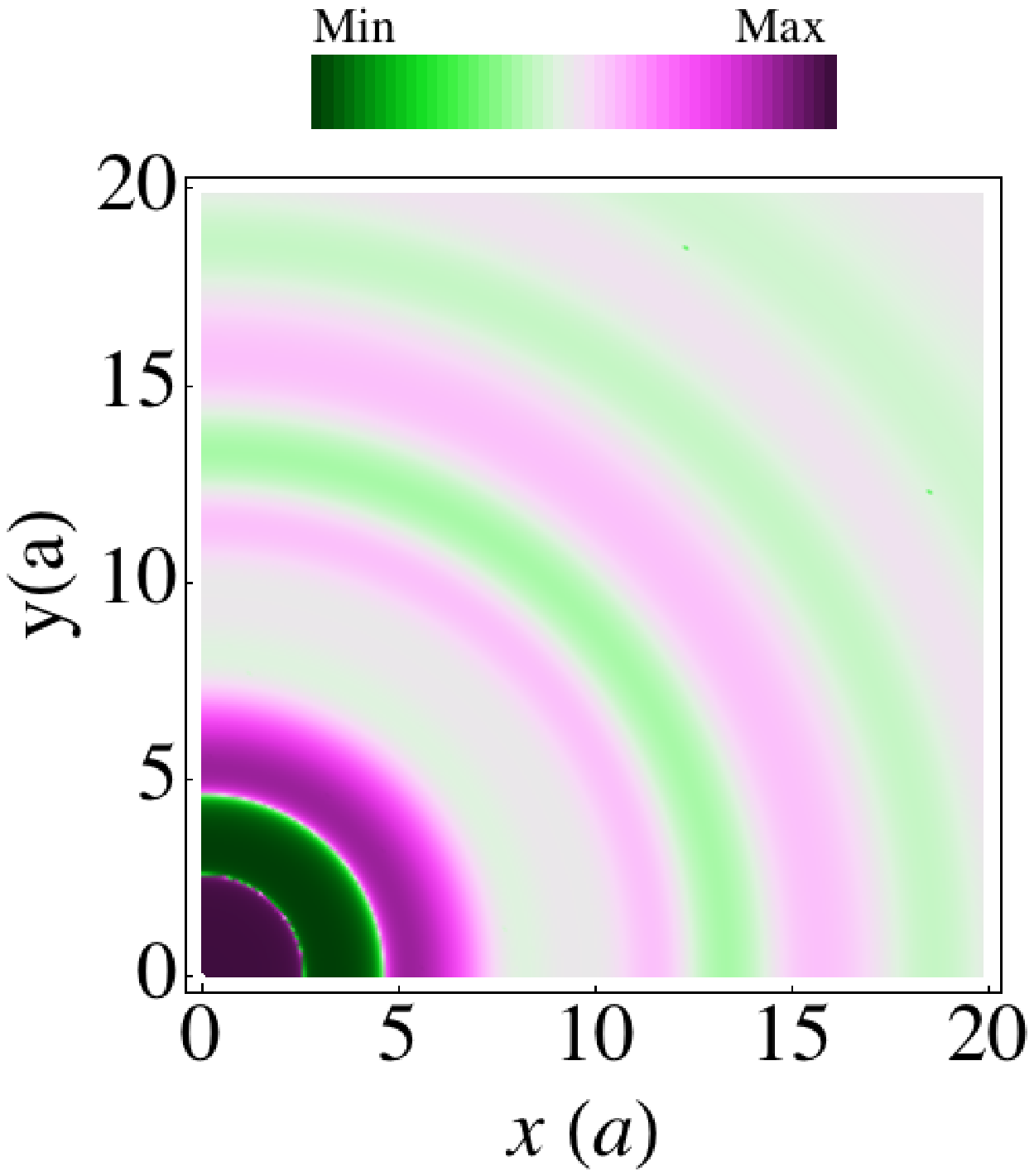}
    }
    \hspace{0.1cm}
            \subfigure[]
        {
        \includegraphics[width= 0.27\linewidth]{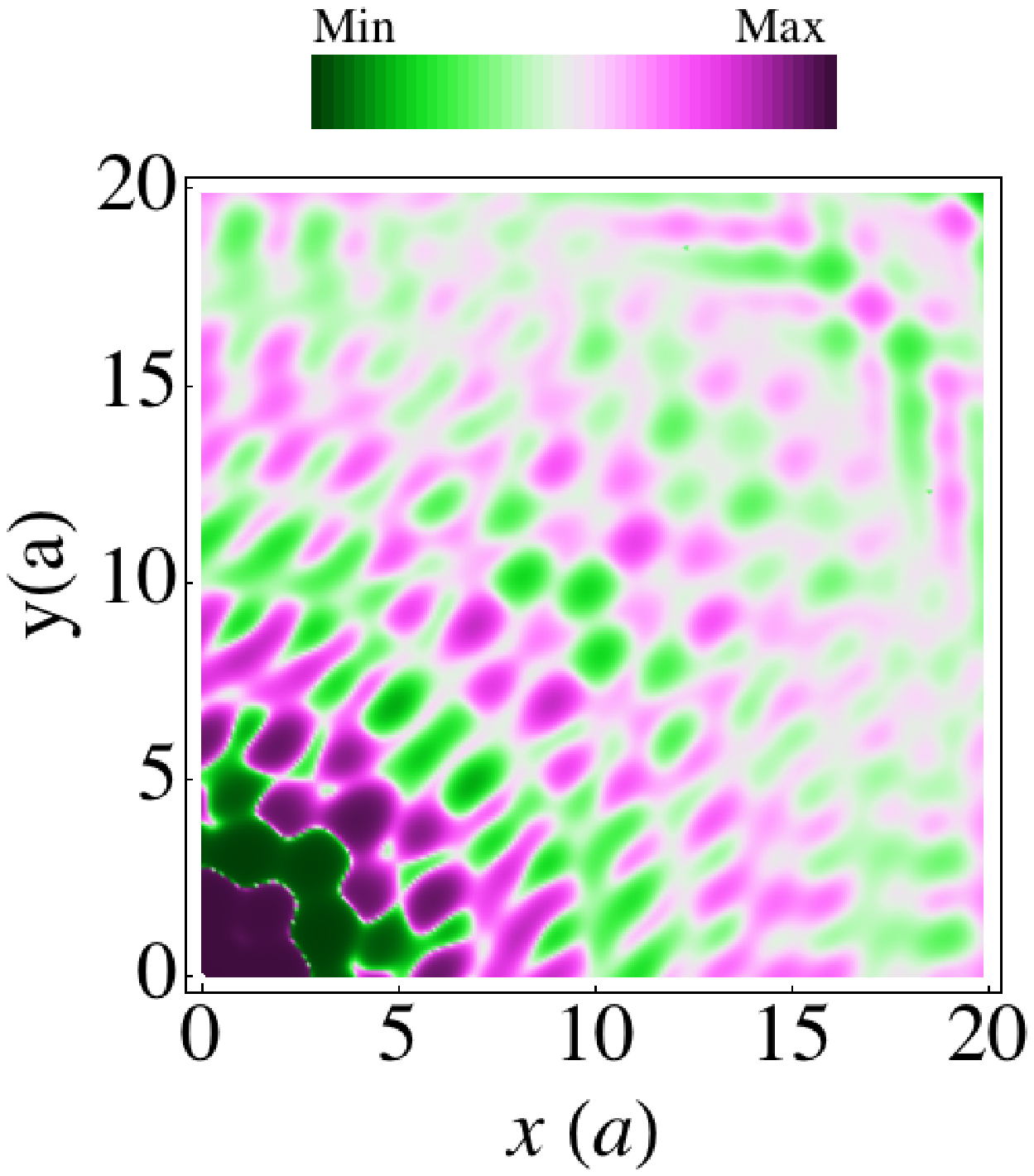}
    }
   \\\vspace{0.4cm} \hspace{-0.10cm}
        {
        \includegraphics[width= 0.85\linewidth]{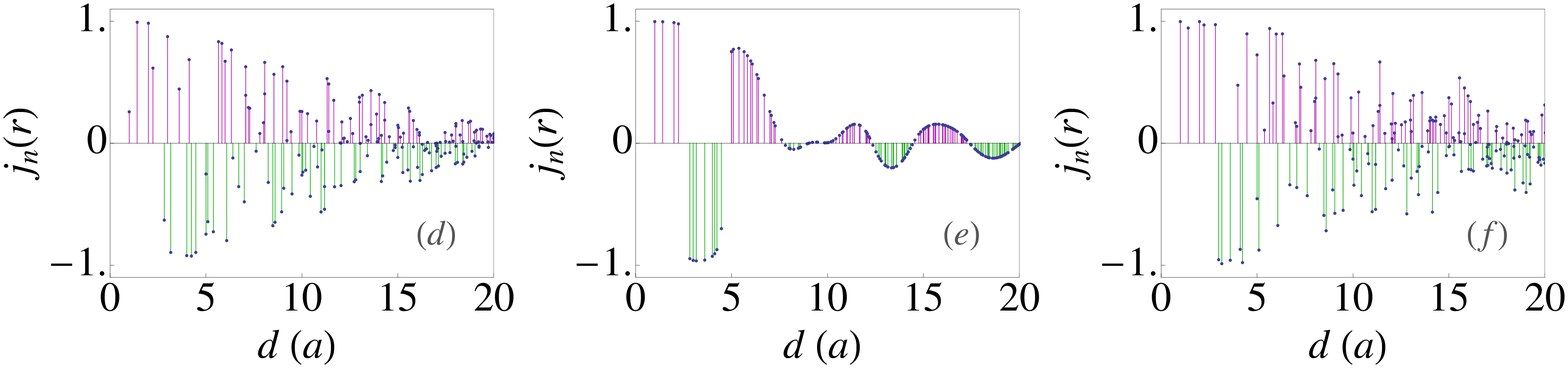}
    }
\caption{(Color online)
 The spatial plot of the  normalized total RKKY  interaction,  $j_n({\bf r})=\frac{J_n({\bf r})}{| J_n({\bf r})|+0.015}$, for a) pure  electron- like pockets,  b) pure hole- like pockets and c) for the case with 4 electron and hole bands, corresponding to  Fig.(\ref{fig1}).\break
Bar plots:   (d), (e) and (f) present  of normalized total RKKY  interaction, $\langle j_n({\bf r}) \rangle$, for two local moment sitting on the origin and the lattice site $(x,y)=(na,ma); n,m =0,\pm1 ..$ with distance $d=\sqrt{x^2+y^2}$ corresponding to the plots   (a), (b) and (c), respectively.
 \vspace{0.4cm}
 }
\label{fig4}
\end{figure}
%

\section{Model definition }
\label{sect:model}

The Hamiltonian describing localized magnetic moments 
in the multi-band conduction electron sea is given by
\bea
{\cal H}={\cal H}_c  + {\cal H}_{int},
\eea
where ${\cal H}_c$ is the conduction electron Hamiltonian according to
\be
{\cal H}_c
= \sum_{{\bf  k}, \gamma, \sigma}
\varepsilon^{\gamma}_{{\bf  k}}
c_{\gamma{\bf  k}\sigma}^\dag c_{\gamma{\bf  k} \sigma}.
\label{Hamiltonian}
\ee
Here, $c_{\gamma{\bf  k} \sigma}^\dagger$ denotes the creation operators and
$\varepsilon^{\gamma}_{{\bf  k}}$ is the dispersion
of conduction electrons with momentum ${\bf k}=(k_x,k_y)$ and  spin $\sigma$  in band $\gamma$.

In the present work we  consider primarily 
the minimal four-bands model consisting two hole- like pockets centered at $\Gamma$- point  ($\alpha$-bands), and  two elliptical electron FS pockets centered  around $\mathbf{Q}_\beta$ [$\mathbf{Q}_1=(\pm \pi, 0)$ and $\mathbf{Q}_2=(0, \pm \pi)$] points in the unfolded Brillouin zone (BZ) ($\beta$-bands). We consider three different cases: a) only electron- like FS sheets, b) only hole- like FS sheets, and finally   c) both hole and electron FS sheets  (see Fig.\ref{fig1}). Thus one can rewrite the total  conduction electron Hamiltonian Eq.(\ref{Hamiltonian}) as 
\bea
{\cal H}_c
=
 \sum_{\alpha,{\bf  k}, \sigma}
\varepsilon^{\alpha}_{{\bf  k}} a_{\alpha{\bf  k} \sigma}^\dag a_{\alpha{\bf  k} \sigma} +
 \sum_{\beta,{\bf  k}, \sigma}\varepsilon^{\beta}_{{\bf  k}} b_{\beta{\bf  k} \sigma}^\dag b_{\beta{\bf  k} \sigma} ,
\eea
 where $a_{\gamma{\bf  k} \sigma}^\dag$ ($b_{\gamma{\bf  k} \sigma}^\dag$) creates electrons  with spin $\sigma $, momentum ${\bf  k}$ at hole-(electron-) like band $\alpha$ ($\beta$). The dispersion of the electrons pockets  can be modeled  as
$
\varepsilon^{\beta}_{\mathbf{k}-\mathbf{Q}_\beta}=\frac{\hbar^2k_x^2}{m^{\beta}_x}+\frac{\hbar^2k_y^2}{m^{\beta}_{y}}-\mu_\beta$,
and for the hole- like pockets we have 
$\varepsilon^{ \alpha}_{\mathbf{k}}=-[\frac{\hbar^2k_x^2}{m^{\alpha}_x}+\frac{\hbar^2k_y^2}{m^{\alpha}_{y}}]-\mu_\alpha$,
where $m^\gamma_n$ denotes the band mass ($n=x,y$) in $k_n$ direction. Furthermore we define  $\mu_\beta=\mu+\varepsilon_0$, and $\mu_\alpha=\mu-\varepsilon_0$ where $\varepsilon_0$ and  $\mu$ are the energy offset and the chemical potential, respectively.
 We consider for the hole- like pockets $m^\alpha_n=m^\alpha$ with $m^{a_2}=3m^{a_{1}}=3m_e$,
  and for the electron- like pockets $m^{b_1}_x=m^{b_2}_y=(1+ \epsilon)m_e$,
    $m^{b_1}_y=m^{b_2}_x=(1- \epsilon)m_e$  with ellipticity $\epsilon$ and electron mass $m_e$. The corresponding general FS is shown in Fig.(\ref{fig1}.c).

The interaction  between  the spin of
conduction electrons, ${\bf s}({\bf R})$, and
the moment of localized $f$-electrons, ${\bf S}_i$, at site ${\bf R}_i$  is given by 
\be
{\cal H}_{int}=J_{ex}\sum\limits_{{\bf R}; i}  I({\bf R}-{\bf R}_i)\; {\bf s}({\bf R}) \cdot {\bf S}_i,
\ee
where the exchange integral $ I({\bf R}-{\bf R}_i)$ is established  by the overlap of 3d conduction and localized 4f electron wave functions. 
The real overlap of the conduction electron and localized 4f electron wave-functions and therefore  $ I({\bf R}-{\bf R}_i)$   in general is non-zero in a finite volume around the 4f site $\bR_i$. This leads to a damping in the effective intersite coupling energies
\cite{Szalowski:2008,Smirnov:2009,Smirnov:2010}. For simplicity we use here the common approximation of an on-site exchange interaction which can be approximated as   $ I({\bf R}-{\bf R}_i)=J_{ex}\; \delta({\bf R}-{\bf R}_i)$.
Here $J_{ex}$ is the on-site exchange coupling constant that can be obtained from a more microscopic Anderson-type model \cite{Akbari:2010lr} by a Schrieffer-Wolff transformation.

\section{Spatial and momentum dependence of effective RKKY interaction}
\label{sect:RKKY}

The contact exchange interaction with local moments will polarize the conduction states which then
leads to an effective RKKY exchange between moments at different sites. It may be obtained by using
the standard second-order perturbation theory with respect to ${\cal H}_{int}$.
Here we consider first the general form of the RKKY interaction of a pair of local moments and its distance 
dependence for the various FS topologies. 
In addition we investigate the momentum dependence of the effective RKKY exchange which determines the 
magnetic structure in a periodic lattice of local moments.

\subsection{Localized magnetic  moment pairs  and the distance dependence of RKKY interaction}
\label{subsect:RKKY_pair}

In the paramagnetic or normal state regime the  effective exchange Hamiltonian of the RKKY interaction which describes the interaction
between two local impurity spins at the positions $i$ and $j$ is derived as \cite{Aristov:1997qy,Akbari:2011fk}
\be
{\cal H}_{RKKY}^{  ij}=-J_n({\bf r}){\bf S}_{i}\cdot {\bf S}_{j},
\ee
where ${\bf r}={\bf R}_i-{\bf R}_j=x{\hat{\it {\bf  x}}}+y {\hat{\it {\bf y}}}$, and the effective exchange couplings are then given by
\be
J_n({\bf r})=\sum\limits_{\gamma\gamma^\prime} {\cal J}^{  \gamma\gamma^\prime}_{n}({\bf r})=J_{ex}^2
Re\left(\sum\limits_{\gamma\gamma^\prime}
e^{i({\bf Q}_{\gamma}-{\bf Q}_{\gamma^\prime})\cdot {\bf r}}  \chi^{\gamma\gamma^\prime} ({\bf r})\right),
\ee
here
$\chi^{\gamma\gamma^\prime} ({\bf r})$ denotes the intra- ($\gamma=\gamma'$) and inter- ($\gamma\neq\gamma'$)   magnetic spin susceptibility of conduction electrons (Lindhard response function) which is defined by
\be
\chi^{\gamma\gamma^\prime} ({\bf r})=-
k_BT\sum\limits_{n}
G_{\gamma}({\bf r},i\omega_n)G_{\gamma^\prime}({\bf r},i\omega_n).
\label{susceptibility_def}
\ee
Here $\omega_n=\pi T (2n+1)$ is the fermionic Matsubara frequency and 
\be 
G_{\gamma}({\bf k},i\omega_n)=
\frac{1}{i\omega_n-\varepsilon^{\gamma}_{\mathbf{k}}},
\ee
is  the conduction electrons GreenÕs function in momentum representation.
After some algebra \cite{Aristov:1997ly} the real space GreenÕs function can be obtained  as
\be
G_{\gamma}({\bf r},i\omega_n)=\frac{\sqrt{m_x^\gamma m_y^\gamma}}{\pi\hbar^2}K_0(\sqrt{2Z_{\gamma}(\omega_n)\rho_\gamma}),
\label{green_real}
\ee
where $K_0(\ldots)$ is the modified Bessel (Macdonald) function.
Here we define 
$Z_{\gamma}=\mu_{\gamma}+i\omega_n$ and $\rho_{\gamma}=(m_x^{\gamma}x^2+m_y^{\gamma} y^2)/\hbar^2$.
In the   low temperature regime  
$$k_BT\sum_{ n} (\ldots)\longrightarrow \frac{1}{2\pi i}\int_{-i \infty}^{i\infty}d\omega (\ldots),$$
therefore
the magnetic spin susceptibility, Eq.(\ref{susceptibility_def}), can be calculated using the
Green's function
as~\cite{Schwabe:1996uq}
\be
\chi^{\gamma\gamma^\prime} ({\bf r})
=
\frac{1}{2\pi i}
\int_{-i \infty}^{i\infty}
d\omega
G_{\gamma}({\bf r},i\omega)G_{\gamma^\prime}({\bf r},i\omega).
\label{susceptibility}
\ee
%


The closed analytic expression for the above integral may be obtained in the case of having the same kind of bands,   where  $\mu_\gamma=\mu_{\gamma^\prime}$ (Fig.\ref{fig1}.a or Fig.\ref{fig1}.b).
Explicitly, using the properties of the modified Bessel functions  in the limit  of $Z_\gamma \to \mu_\gamma+i0^{\pm}$, we have 
\be
K_n (\sqrt{2Z_e \rho_{\gamma}}) = \frac{\pm \pi i}{2}   e^{\pm n \pi i/2} H_n^{(\pm)}(\sqrt{2\mu_\gamma\rho_{\gamma}}), 
\ee
where $H_n^{(\pm)}(x)={\mbox J}_n(x)\pm i {\mbox Y}_n(x) $ are Hankel functions \cite{Abramowitz:1984}.

In the special case  $\rho_\gamma=\rho_{\gamma'}$, by using Eqs.~(\ref{green_real},\ref{susceptibility}), the magnetic susceptibility   can be written  as
\be
\chi^{\gamma\gamma^\prime} ({\bf r})
=
\Phi_{\gamma\gamma^\prime}
\Bigg[
{\mbox J}_0(X_{\gamma}) {\mbox Y}_0(X_{\gamma}) +{\mbox J}_1(X_{\gamma}) {\mbox Y}_1(X_{\gamma}) 
\Bigg],
\label{chi_gamma_gamma}
\ee
here $\Phi_{\gamma\gamma^\prime}=-\sqrt{m_x^{\gamma} m_y^{\gamma} m_x^{\gamma^\prime} m_y^{\gamma^\prime}}
 \frac{
 \mu_{\gamma} }{4\pi \hbar^4}$ and $X_{\gamma}=\sqrt{|2\mu_{\gamma}\rho_{\gamma}|}$.

In the general case $\rho_\gamma\neq\rho_{\gamma'}$ we still can find the closed solution
\bea
\chi^{\gamma\gamma^\prime} ({\bf r})
&=&
\frac{\Phi_{\gamma\gamma^\prime}
}{X^2_\gamma-X^2_{\gamma'}}
\nonumber\\
&&\times
{\Bigg [  }
X_{\gamma}
{\Big (  }
{\mbox J}_1(X_{\gamma}){\mbox Y}_0(X_{\gamma'})+{\mbox Y}_1(X_{\gamma}){\mbox J}_0(X_{\gamma'})   
{\Big ) }
\nonumber\\
&&
-X_{\gamma'}
{\Big (  }
{\mbox J}_1(X_{\gamma'}){\mbox Y}_0(X_{\gamma}) +{\mbox Y}_1(X_{\gamma'}){\mbox J}_0(X_{\gamma})    
{\Big )  }
{\Bigg ]}. 
\nonumber\\
\label{chi_gamma_gammaP}
\eea

For the simple case of  spherical FS sheets, i.e.   $m_x^\gamma=m_y^\gamma$ we have  $X_{\gamma}=k_f^\gamma r$,
where $k_f^\gamma$ is the Fermi momentum, and $r=\sqrt{x^2+y^2}$ the distance between local moments.

%
\begin{figure}
 \centering
         \subfigure[]
    {
        \includegraphics[width= 0.3\linewidth]{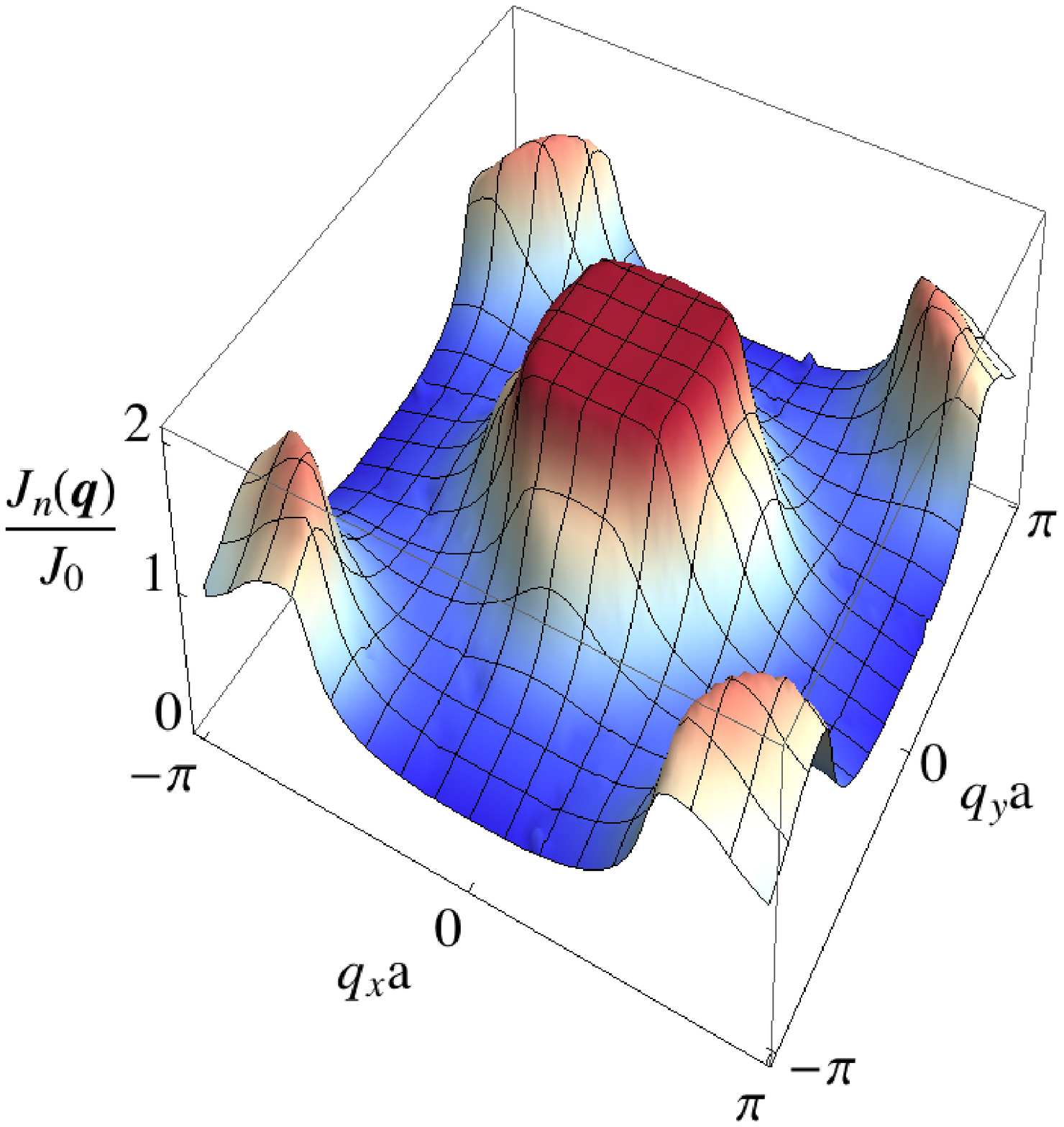}
    }
    \hspace{0.1cm}
    \subfigure[]
    {
        \includegraphics[width= 0.3\linewidth]{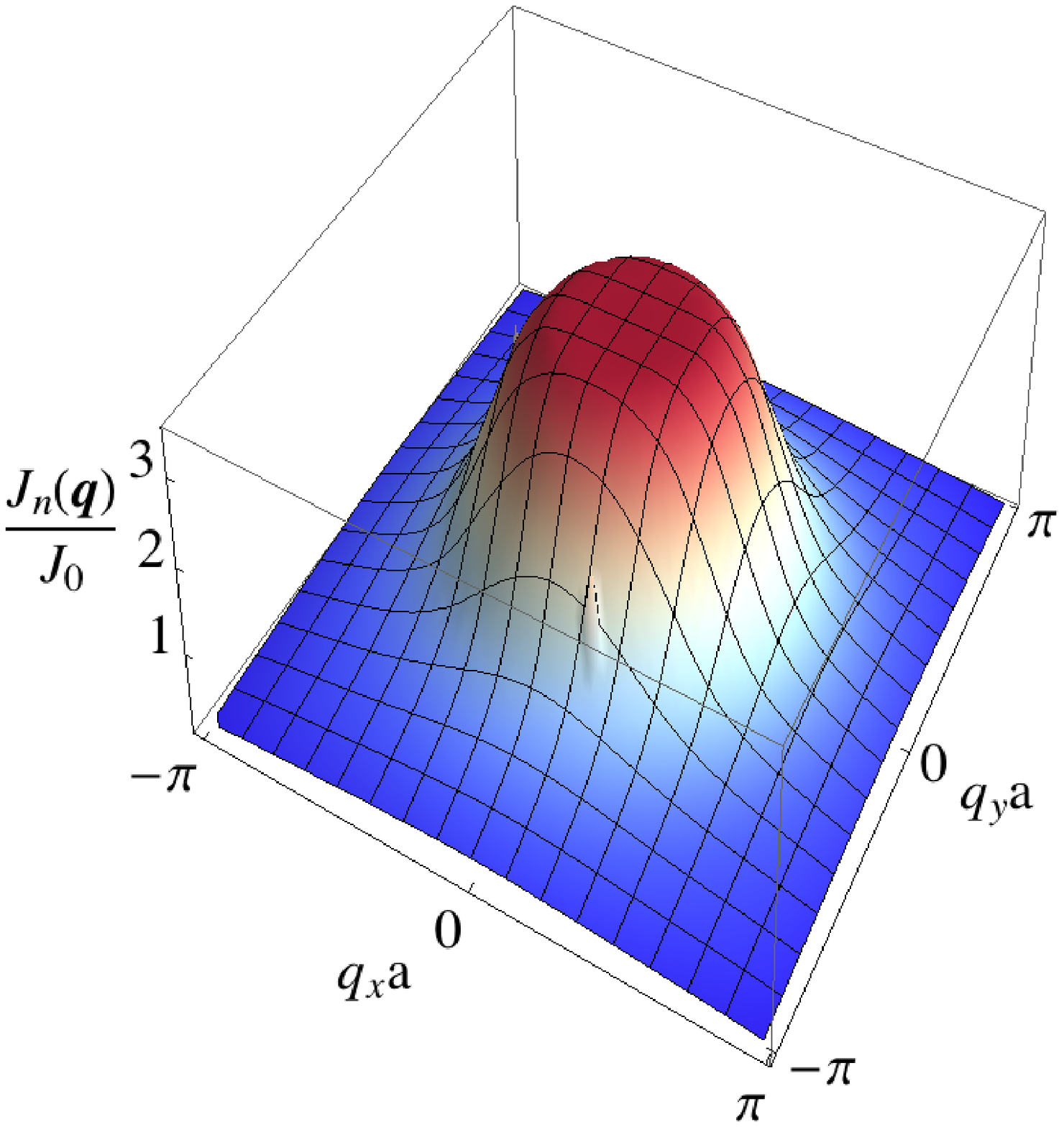}
    }
   \\
    \subfigure[]
        {
        \includegraphics[width= 0.3\linewidth]{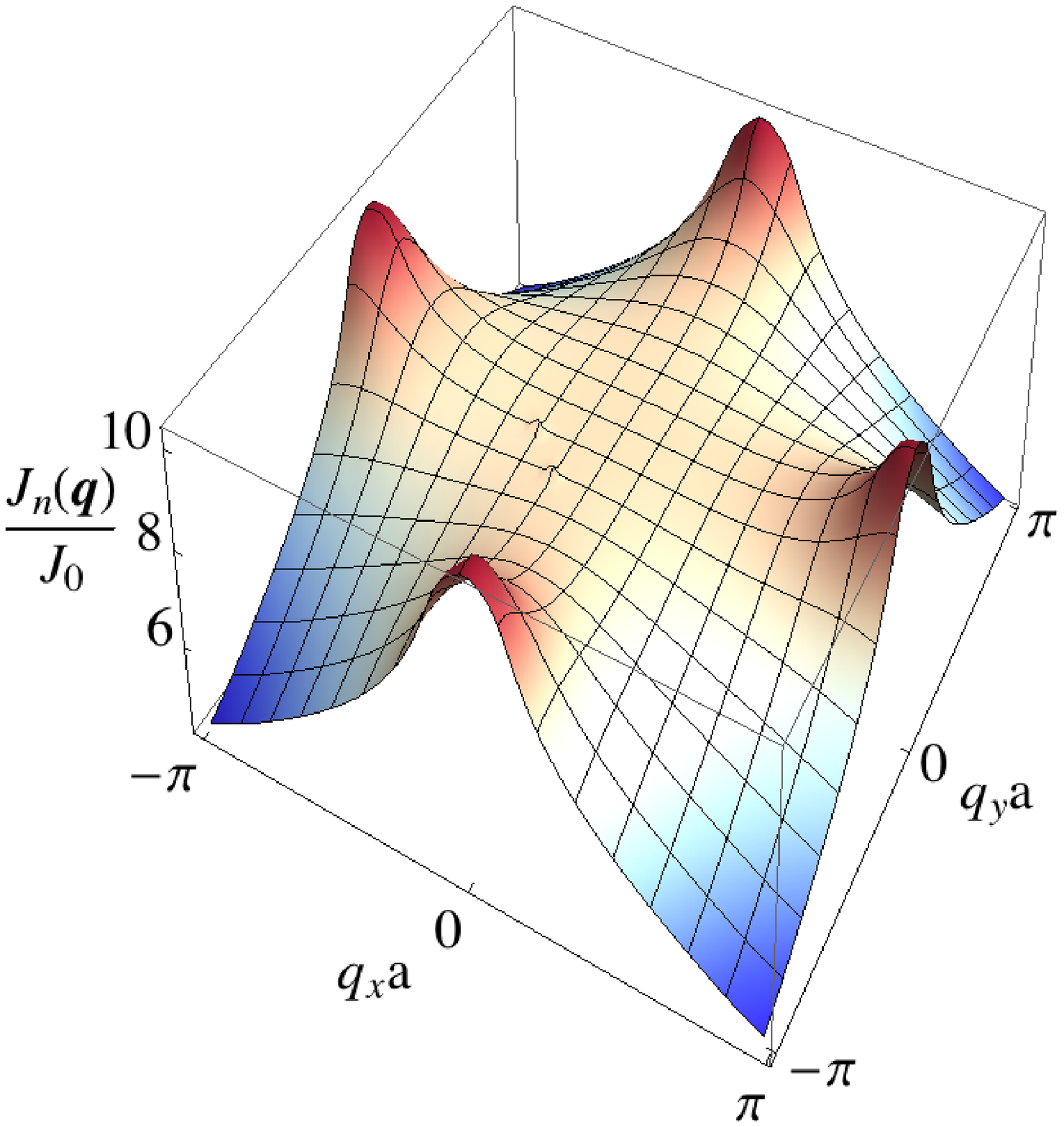}
    }
    \hspace{0.1cm}
    \subfigure[]
        {
        \includegraphics[width= 0.3\linewidth]{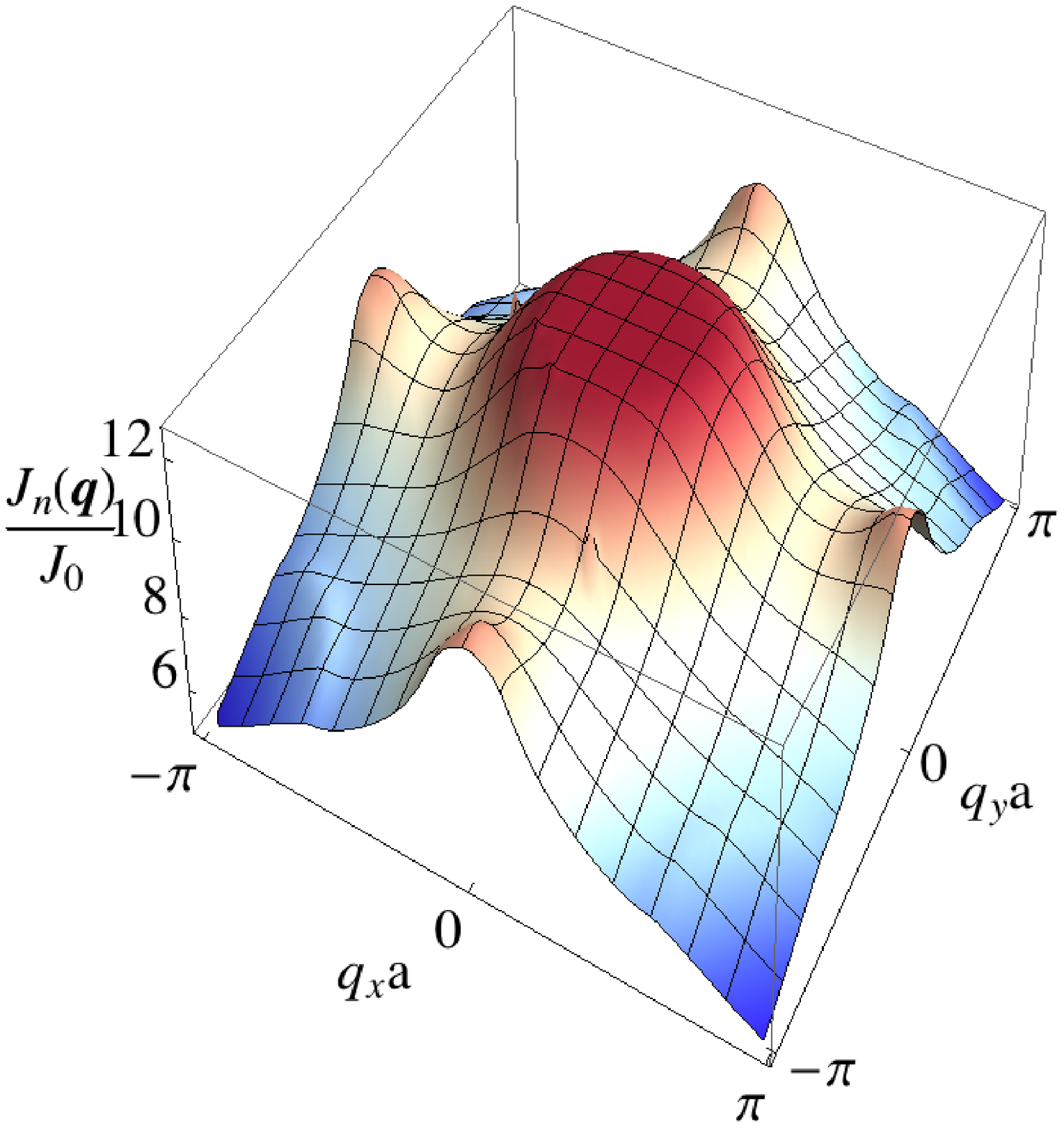}
    }
\caption{(Color online)
 The momentum dependency of the Fourier transformation of the  total RKKY interaction,  $J_n({\bf q})$  for a) pure  electron- like pockets,  b) pure hole- like pockets  c)  only the inter hole- and electron- like contributions and d) the to total interaction corresponding to sum of all FS sheet contributions. Results are obtained for the simple parabolic  dispersion corresponding to Fig.(\ref{fig1}). Absolute maximum of   $J_n({\bf q})$ in this simplified model occurs at ${\bf q}=0$ and side maxima at ${\bf q}={\bf Q}_\alpha$.
 \vspace{0.4cm}
 }
\label{fig5}
\end{figure}
%
%
\begin{figure}
 \centering
          \subfigure[]
          {
\includegraphics[width=0.44 \textwidth]{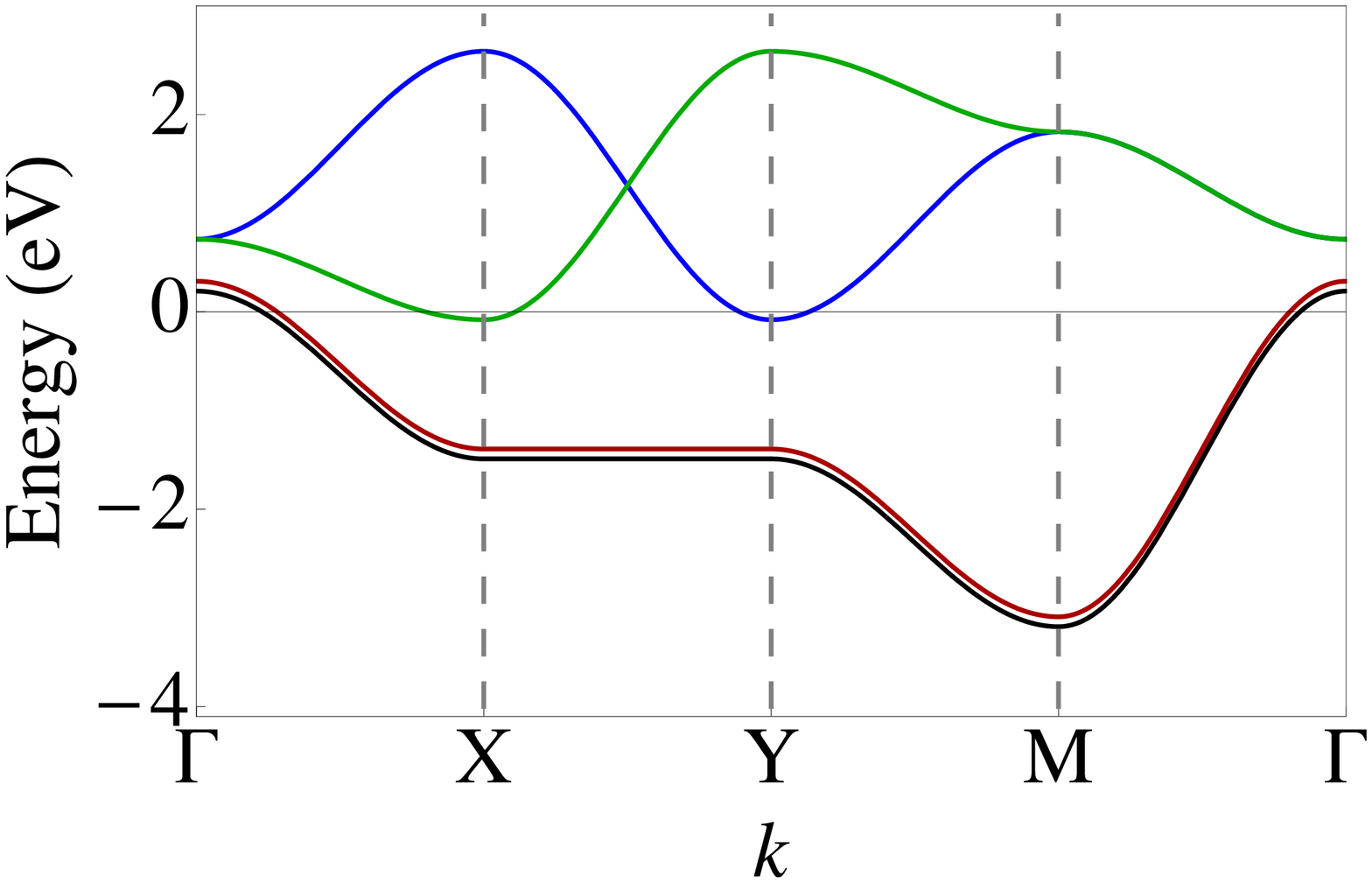}
}
         \subfigure[]
         {
\includegraphics[width=0.421 \textwidth]{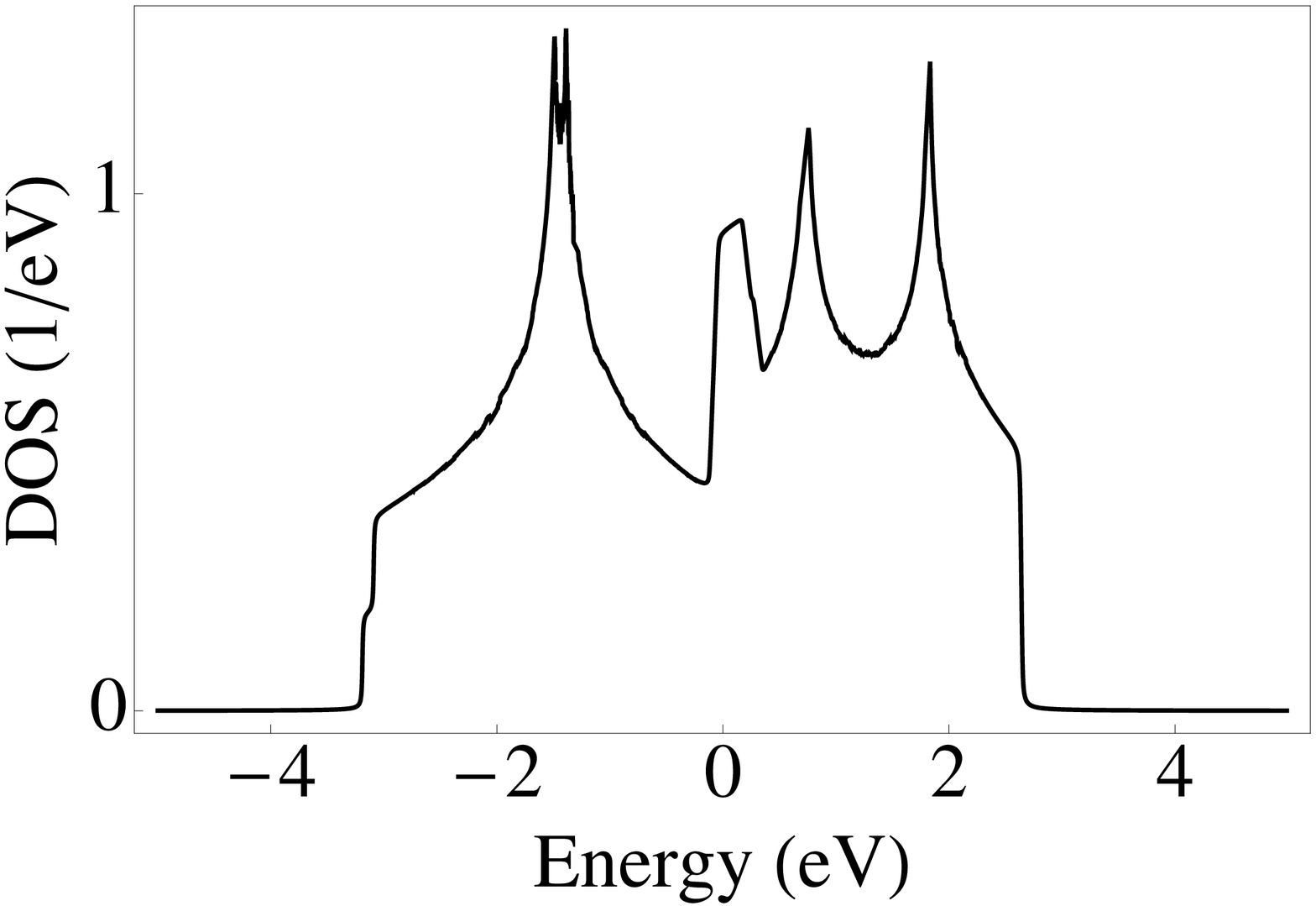}
}
    \\
      \hspace{-0.25cm}
         \subfigure[]
    {
        \includegraphics[width=0.3\linewidth]{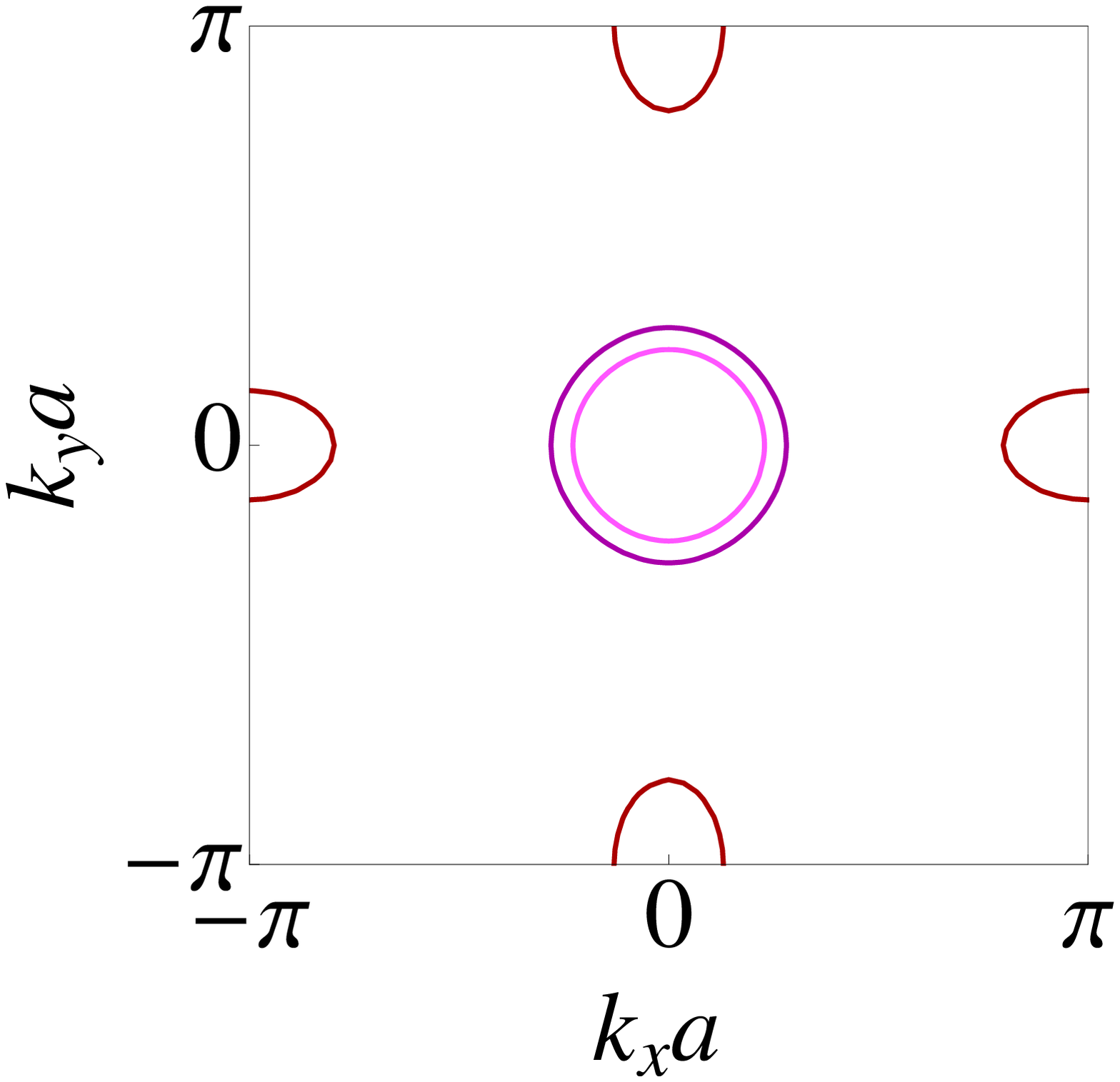}
    }
            \hspace{0.25cm}
             \subfigure[]
    {
        \includegraphics[width=0.3\linewidth]{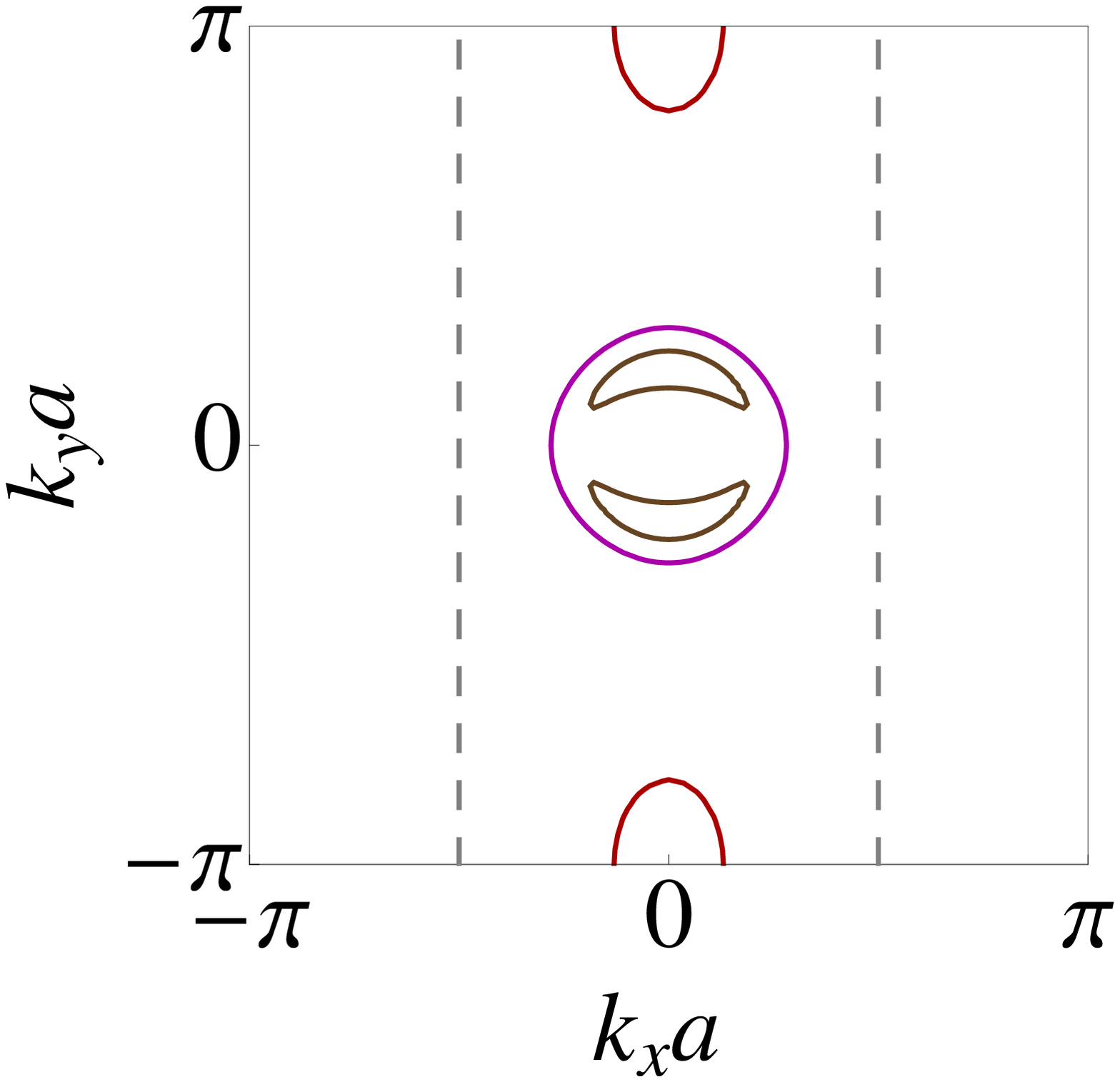}
    }
\caption{(Color online)
(a) A typical tight binding band structure for iron based  122-type superconductors with  two circular  hole pockets and two  electron- like  pockets.
The BZ symmetry points are $\Gamma (0,0)$, $X(0,\frac{\pi}{a})$, $Y(\frac{\pi}{a},0)$ and $M(\frac{\pi}{a},\frac{\pi}{a})$.
The band structure parametrization can be found in Ref.\cite{Akbari:2010}.
(b) The corresponding  density of states (DOS), and 
(c) its Fermi surface. 
(d) The reconstructed FS in the SDW phase with ordering parameter $W=0.15 eV$.
}
\label{fig6}
\end{figure}
%

\subsection{Periodic lattice of local moments  and the momentum dependence of RKKY interaction}
\label{subsect:RKKY_lattice}

In the case  of local moments forming a regular lattice it is necessary to go to the 
momentum representation of the RKKY interaction which is defined by
\be
{\cal H}_{ff} =
-\sum_{ij}J_n({\bf r}){\bf S}_{i}\cdot {\bf S}_{j}
=-\sum_{\bf q}J_n({\bf q}){\bf S}_{-{\bf q}}\cdot {\bf S}_{{\bf q}},
\ee
where the real space RKKY interaction is given by
\be
J_n({\bf r})=J_{ex}^2\sum\limits_{\gamma\gamma^\prime}
e^{ i({\bf Q}_\gamma-{\bf Q}_{\gamma'})\cdot {\bf r}}
\chi^{\gamma\gamma^\prime} ({\bf r}),
\label{RKKYr}
\ee
and
\be
J_n({\bf q})=J_{ex}^2\sum\limits_{\gamma\gamma^\prime}
\chi^{\gamma\gamma^\prime} ({\bf q}-({\bf Q}_\gamma-{\bf Q}_{\gamma'})),
\label{RKKYq}
\ee  
is its Fourier transform and 
${\bf S}_{{\bf q}}$ are the Fourier components of ${\bf S}_i$.
They are defined by 
\be
J_n({\bf q})=\frac{1}{\sqrt{N}}\sum_{\bf r}J_n({\bf r})e^{i {\bf q}\cdot {\bf r}}, \;\;
S_{\bf q}=\frac{1}{\sqrt{N}}\sum_{i}S_ie^{i {\bf q}\cdot {\bf R}_i},
\ee
respectively, with $N$ denoting the lattice size.
The RKKY interaction is proportional to the static conduction electron magnetic susceptibility given by
\be 
 \chi^{\gamma\gamma^\prime} ({\bf q})
=
-\frac{1}{N}\sum_{\bf k}
\frac{f(\varepsilon^{\gamma}_{\mathbf{k}-{\bf q}})-f(\varepsilon^{\gamma^\prime}_{\mathbf{k}})}
{\varepsilon^{\gamma}_{\mathbf{k}-{\bf q}}-\varepsilon^{\gamma^\prime}_{\mathbf{k}} },
\label{susfin}
\ee
where  $f(\ldots)$ denotes the Fermi function. Here we implied that the electron pocket has been shifted to
the $\Gamma $- center of the BZ. We may also conveniently define a total susceptibility 
\be
\chi_t({\bf q})=\sum\limits_{\gamma\gamma^\prime}
\chi^{\gamma\gamma^\prime} ({\bf q}-({\bf Q}_\gamma-{\bf Q}_{\gamma'})).
\label{sustot}
\ee  
The meaning of $\chi_t({\bf q})$ is twofold: At the wave vector {\bf q} where $\chi_t({\bf q})$ acquires its maximum an SDW instability may occur due to 3d conduction electron interactions, opening a SDW gap at FS patches connected by this momentum. Furthermore at the same wave vector the ordering of 4f local moments due to $J_n({\bf q})=J_{ex}^2\chi_t({\bf q})$ in the rare earth - based Fe pnictides should be expected at lower temperature. Note, however, that due to the feedback effect the opening of the SDW gap may
influence the wave vector where $J_n({\bf q})$ has its maximum.

%
\begin{figure}
 \centering
         \subfigure[]
    {
        \includegraphics[width= 0.3\linewidth]{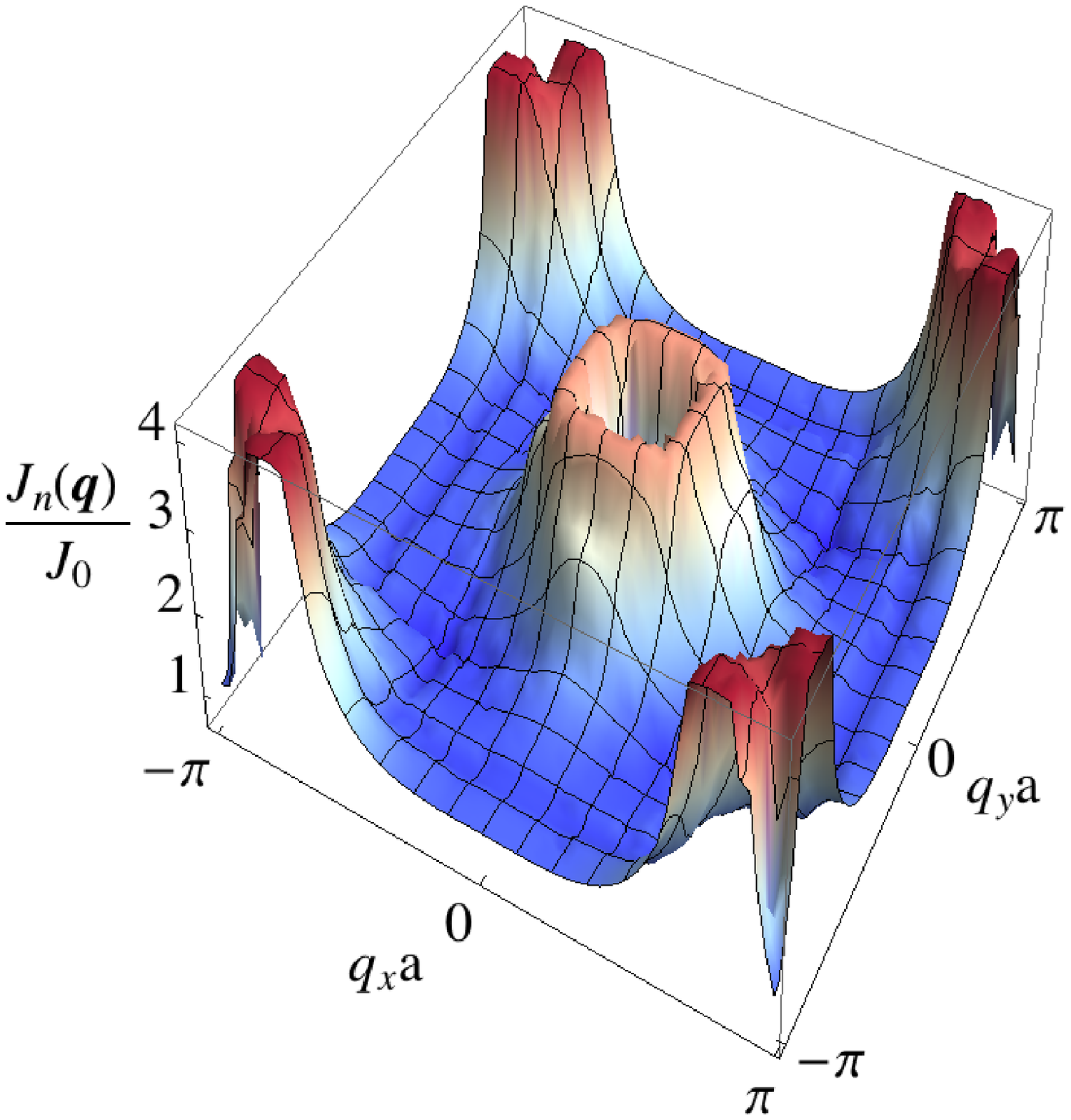}
    }
    \hspace{-0.1cm}
    \subfigure[]
    {
        \includegraphics[width= 0.3\linewidth]{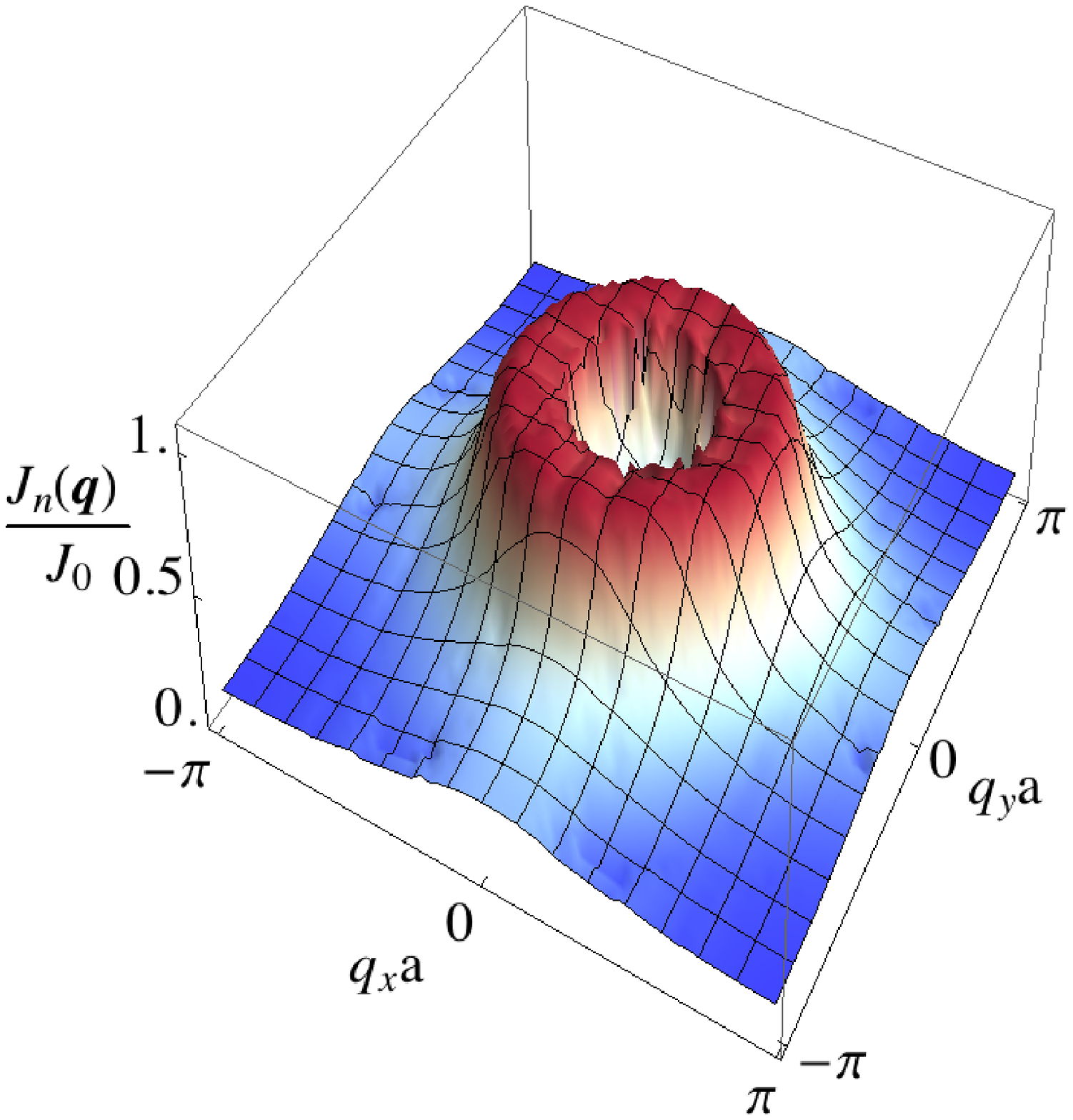}
    }
    \\
    \subfigure[]
        {
        \includegraphics[width= 0.3\linewidth]{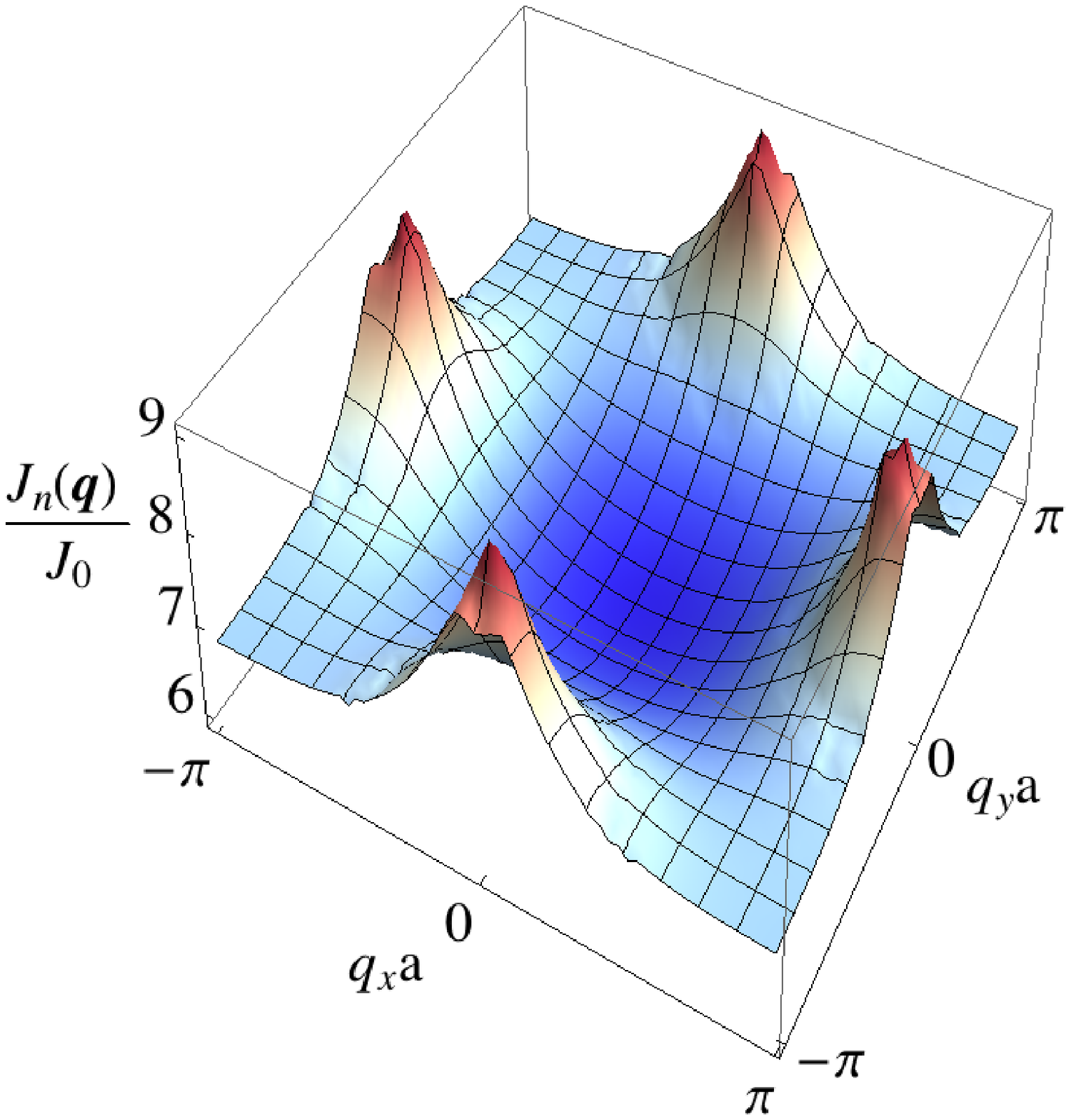}
    }
    \hspace{-0.1cm}
    \subfigure[]
        {
        \includegraphics[width= 0.3\linewidth]{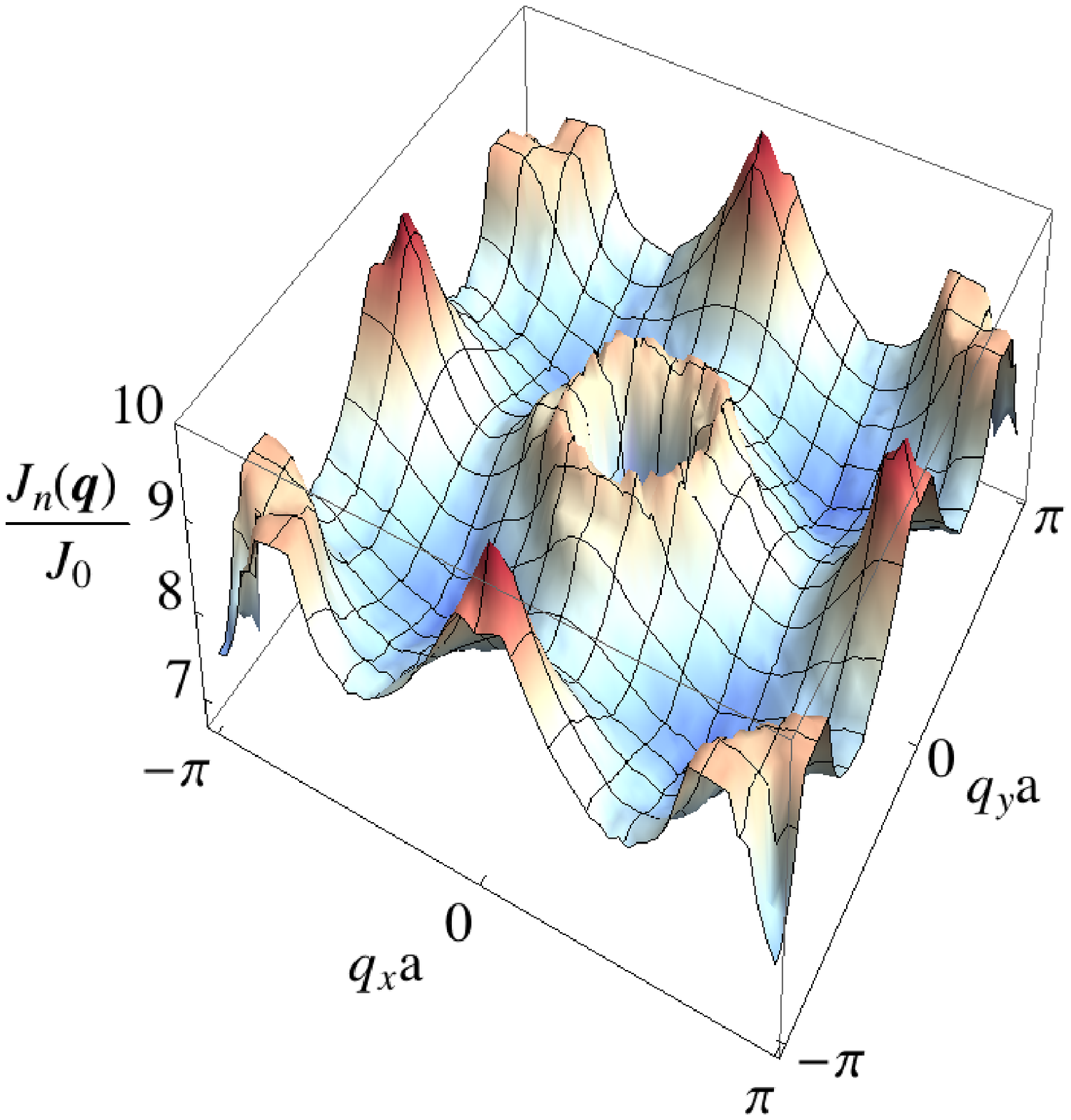}
    }
\caption{(Color online)
 The momentum dependence of the Fourier transform of the  total RKKY interaction,  $J_n({\bf q})$, for a) pure  electron- like pockets,  b) pure hole- like pockets, c) only the inter hole- and electron- like contributions and d) the total interaction corresponding to sum of all  contributions. 
Results correspond to a typical tight-binding type band structure  example (see Fig.(\ref{fig6})) for 122 compounds.
 \vspace{0.4cm}
 }
\label{fig7}
\end{figure}
%
%

\subsection{Momentum dependence of RKKY interaction in spin density wave phase}
\label{subsect:RKKY_latticeSDW}
Recently it was shown \cite{Akbari:2011fk} that
the change of the FS nesting (resulting of the opening of the SDW gap)  influences both  strength and oscillatory behavior of RKKY in the antiferromagnetic state.
In the mean-field approximation the SDW ordering can be described by \cite{Eremin:2010lr} 
\be
{\cal H}_{SDW}^{MF}=\sum\limits_{{\bf k} \sigma}
W \sigma
\left[ a^\dag_{1{\bf  k}\sigma} b_{1{\bf k}+{\bf Q}_1 \sigma}+ H.c.\right],
\ee
where $W$ is the ordering amplitude  and the spin index $\sigma=\pm$ corresponds to spin $\uparrow$ and $\downarrow$.
The SDW ordering  induces an anisotropy in the RKKY interaction 
and   maps the RKKY into an effective anisotropic XXZ-type Heisenberg exchange model,
\bea
{\cal H}_{ff} 
&=&
-\sum_{ij}{\Bigg [  }J_x({\bf r}){\Big (  }{\bf S}_{i}^x{\bf S}_{j}^x+{\bf S}_{i}^y{\bf S}_{j}^y{\Big ) }+J_z({\bf r}){\bf S}_{i}^z{\bf S}_{j}^z
{\Bigg ]  }
\nonumber
\\
&=&
-\sum_{q}{\Bigg [  }J_x({\bf q}){\Big (  }{\bf S}_{{\bf q}}^x{\bf S}_{-{\bf q}}^x+{\bf S}_{{\bf q}}^y{\bf S}_{-{\bf q}}^y{\Big ) }+J_z({\bf q}){\bf S}_{{\bf q}}^z{\bf S}_{-{\bf q}}^z
{\Bigg ]  },
\nonumber
\\
\eea 
where the momentum dependence of the Fourier transform of the  total RKKY interaction in SDW phase is obtained as
\bea
&&J_{x,z}({\bf q})=
J_{ex}^2\sum\limits_{\gamma\gamma^\prime ,{\bf k}}
\Psi^{x,z}_{\gamma\gamma^\prime {\bf k} {\bf q}}
\frac{f(E^{\gamma}_{\mathbf{k}-{\bf q}})-f(E^{\gamma^\prime}_{\mathbf{k}})}
{E^{\gamma}_{\mathbf{k}-{\bf q}}-E^{\gamma^\prime}_{\mathbf{k}} },
\label{RKKY_q_SDW}
\eea
with the quasiparticle energies  defined by
$
 E^{ 3}_{{\bf  k}}= \varepsilon^{b_{2}}_{{\bf  k}}$, 
 $ E^{ 4}_{{\bf  k}}= \varepsilon^{a_{2}}_{{\bf  k}}$, and
$$
E^{ 1,2}_{{\bf  k}}=\frac{1}{2}\left[
(\varepsilon^{a_1}_{{\bf  k}}+\varepsilon^{b_1}_{{\bf  k}})\pm \sqrt{
(\varepsilon^{a_1}_{{\bf  k}}-\varepsilon^{b_1}_{{\bf  k}})^2+4W^2 }
\right].
$$
The coherence factors $\Psi^{x,z}_{\gamma\gamma^\prime {\bf k} {\bf q}}$ are given by
\be
 \Psi^{x}_{\gamma\gamma^\prime,  {\bf k},{\bf q}}=\Upsilon_{ {\bf k}-{\bf q}, +}^{\gamma}\Upsilon_{ {\bf k}, -}^{\gamma^\prime } 
 +
\Upsilon_{ {\bf k}-{\bf q}, -}^{\gamma}\Upsilon_{ {\bf k},+}^{\gamma^\prime },
\ee
and 
\be
 \Psi^{z}_{\gamma\gamma^\prime,  {\bf k},{\bf q}}=
 \Upsilon_{ {\bf k}-{\bf q},+}^{\gamma}\Upsilon_{ {\bf k},+}^{\gamma^\prime }+ \Upsilon_{ {\bf k}-{\bf q},-}^{\gamma}\Upsilon_{ {\bf k}, -}^{\gamma^\prime } ,
 \ee
with 
$\Upsilon_{ {\bf k},\sigma}^{1,2 }=(u_{\bf k}\pm\sigma v_{\bf k})^2$ and $\Upsilon_{ {\bf k},\sigma}^{3 }=\Upsilon_{ {\bf k},\sigma}^{4 }=1$.
Furthermore 
the coefficients of the unitary transformation are given by
\be  
u_{{\bf k}}^2, v_{{\bf k}}^2=\frac{1}{2}\left[
1
\pm \frac{(\varepsilon^{h}_{{\bf  k}}-\varepsilon^{e_1}_{{\bf  k}})}{\sqrt{
(\varepsilon^{h}_{{\bf  k}}-\varepsilon^{e_1}_{{\bf  k}})^2+4W^2 }
}
\right].
\ee

%

\section{Discussion of numerical results}
\label{sect:numerical}

Now  we  begin our numerical  discussion of the real space and momentum dependence of the RKKY interaction.
For the calculations we use $\mu_\gamma=\hbar^2k_f^{\gamma 2}/2m_e$
with typical values of  Fermi momentum  $k_f^\alpha a=0.15\pi$ for hole ($\alpha$) - like pockets, and  $k_f^\beta a=0.2\pi$ with  ellipticity $\epsilon=0.4$ for electron ($\beta$)- like pockets (Fig.\ref{fig1}.).
Here $a$ is the lattice constant.\\
 
First we present the results for the individual intra- and inter-band contributions to the total spin susceptibility for the various FS models in Fig.(\ref{fig1}), they are shown in  Fig.(\ref{fig2}) in the same sequence. The distance dependence of  individual components of  $\chi^{\gamma\gamma^\prime} ({\bf r})$,  along the x-direction is shown in this figure. We notice that the oscillation pattern is similar for intra-band contributions ($\alpha-\alpha, \beta-\beta$), however their maxima and minima are slightly shifted due to the different FS dimensions visible in Fig.(\ref{fig1}). This is particularly the case in the  inter-band contributions of Fig.(\ref{fig2}.c).
We note that the results  in Figs.(\ref{fig2}.a,b) were obtained from the closed analytical expression in Eq.~(\ref{chi_gamma_gammaP}) which are identical to the numerical results. For the general FS case with both pockets present $\chi^{\gamma\gamma^\prime} ({\bf r})$ can only be calculated numerically.

The total RKKY exchange is proportional to the sum of all ($\gamma\gamma'$) contributions multiplied with a phase factor determined by the pocket distance, i.e., the  nesting vectors ${\bf Q}_\gamma-{\bf Q}_{\gamma^\prime}$. It is shown in  Fig.(\ref{fig3}) for the three Fermi surface cases of  Fig.(\ref{fig1}).
Fig.(\ref{fig3}.a) represents the RKKY interaction for pure electron- like structure.
As a result of the extra inter-band ($\beta-\beta$) phase factors in Eq.~(\ref{RKKYr}), it shows  a new rapid superposed oscillation on top of the fundamental oscillation determined by the FS sheet diameters which is created by summation of individual parts of Fig.(\ref{fig2}.a).

For the second  case with only hole- like pockets, the  total RKKY coupling shows only the oscillations defined by total summation of individual terms. In contrast to the  case of electron pockets the rapid oscillations are  absent because both ($\alpha_1,\alpha_2$) hole pockets are $\Gamma$- centered without a shift between them.

Finally we present  the calculated RKKY interaction for general four band case  along the x-direction in the  Fig.(\ref{fig3}.c). 
In similar way than in the pure electron case, it shows the combination of slow overall and additional rapid oscillation. The latter originate  from both electron-electron ($\beta-\beta$) and electron-hole ($\alpha-\beta$) contribution but not from the hole-hole  ($\alpha-\alpha$) part.  The size of the electronic pockets and the corresponding Fermi momentum  can be  changed  by tuning the chemical potential. Since the oscillation is defined by the $X_{\gamma}(\simeq k_f^\gamma r)$ value,  an increase of $k_f^\gamma$ causes the reduction of the wave length of both oscillation types  and vice versa.\\

For a better understanding of  the RKKY oscillatory behaviour, in Fig.(\ref{fig4}.a-c) we show
 the full spatial dependency of the  normalized total RKKY  interaction,  $j_n({\bf r})=\frac{J_n({\bf r})}{| J_n({\bf r})|+\eta}$ (we use the parameter $\eta=0.015$ for good extremal contrast). Fig.(\ref{fig4}.a) is for pure  electron- like pockets,  Fig.(\ref{fig4}.b) is for pure hole- like pockets and Fig.(\ref{fig4}.c) is for the general case with 4 electron and hole bands.
 For  a FS with only hole pockets (Fig.(\ref{fig4}.b) only long range (radial) oscillations which are almost isotropic appear. When electron pockets are present  (Fig.(\ref{fig4}.a,c) there are superposed short range oscillations (also azimuthal) due to the interband processes. However a shell-like overall structure of FM/AF regions for moderate distances is preserved, with a notable phase shift
by a half period along the (1,1) direction.

 One of the important issues  of the  diluted system with random local moments (e.g. magnetic impurities like $Mn$) sitting on lattice sites is  the conditon for their magnetic ordering. For this purpose it is useful to know the FM/AF oscillatory RKKY behaviour.
 Therefore we present the    normalized total RKKY  interaction,  $ j_n({\bf r}) $, for two local moments sitting at the origin and the lattice site $(x,y)=(na,ma); n,m=0,\pm 1. ..$ with a distance $d=\sqrt{x^2+y^2}$ in the plots of  Fig.(\ref{fig4}.d-f), respectively. 
 These plots clearly show the change of the interaction form AF to FM by varying the distance of the local moment impurities which is much more rapid when electron like pockets are present. The average impurity distance can be controlled by the concentration of the magnetic impurities in the sample\cite{Texier:2012}.\\

An experimental determination of the real space variation of the RKKY interaction is sofar not easily possible. However, its Fourier transform is a more accessible quantity, because the magnetic ordering both in the 3d system and in the effectively RKKY coupled 4f system should occur at wave vectors that are maxima of the total static susceptibility  $\chi_t({\bf q})=J_n({\bf q})/J_{ex}^2$, therefore it is useful to calculate $J_n({\bf q})$. The result for the three different contributions from intra-and inter-band transitions to the susceptibility are shown in  Fig.(\ref{fig5}.a-c). The intra-band contributions have their maximum at the $\Gamma$ point where the electron part (a) has an additional side maximum at the $\beta-\beta$ nesting vector ${\bf Q}=(\pi,\pi)$ which is due to excitations between different electron pockets. The inter-band contribution (c) on the other hand has maxima at
 the ${\bf Q}_\alpha=(\pi,0),(0,\pi)$, i.e.  $\alpha-\beta$ nesting vectors. However the value at the $\Gamma$-point $(0,0)$ is still considerably enhanced. For the general FS of  Fig.(\ref{fig1}.c) these contributions have to be summed up according to Eq.~(\ref{sustot}). The resulting total $\chi_t({\bf q})=J_n({\bf q})/J_{ex}^2$ is shown in  Fig.(\ref{fig5}.d). Obviously the absolute maximum is still at the $\Gamma$ point and strongly peaked side maxima at  ${\bf Q}_\alpha=(\pi,0),(0,\pi)$ are present. This raises a question about the validity of the simple parabolic electron-hole model discussed so far since the magnetic instability in the 3d system of Fe pnictides is of the SDW type with an ordering vector  ${\bf Q}_\alpha=(\pi,0),(0,\pi)$.\\

To understand this issue better we also calculated the momentum dependence of $\chi_t({\bf q})=J_n({\bf q})/J_{ex}^2$ in a more realistic  tight binding (TB) type band structure for the 3d bands valid for the 122 compounds. The Fermi velocities and 
size of the pockets in this model are based on Refs.\cite{Singh:2008,Boeri:2008,Mazin:2008}. The band structure and associated  density of states (DOS) are presented in Fig.(\ref{fig6}). The detailed parametrization  can be found in Ref.\cite{Akbari:2010}.
The basic features of momentum dependence is similar to the previous model: The maxima occur at or close to zone center due to intra-band processes and at zone boundary points due to nesting features of electron-electron (a) and electron-hole (c) excitations. An essential difference to the parabolic pocket model in  Fig.(\ref{fig5}c) is the small value of the electron-hole contribution in  Fig.(\ref{fig7}c) for small momentum transfer. This difference is due to the deep depression in the tight-binding model DOS for $\omega\approx 0$ whereas the DOS for the 2D parabolic band model is simply constant in that region. As a consequence the absolute maximum of  $\chi_t({\bf q})=J_n({\bf q})/J_{ex}^2$  in the tight-binding case is located at the ${\bf Q}_\alpha=(\pi,0),(0,\pi)$ positions ( Fig.\ref{fig7}d).
Therefore in the latter case the SDW instability of the itinerant 3d electrons is predicted at the proper wave vector.

However the simplified model results in  Fig.(\ref{fig5}) are nevertheless useful for the understanding of the RKKY interaction because the latter refer to the ordering of localized 4f electrons which takes place only at temperatures much lower than the SDW transition temperature. In the low temperature region the fully developed SDW gap of 3d band electrons will strongly suppress the peak in 
$\chi_t({\bf q})=J_n({\bf q})/J_{ex}^2$  at  ${\bf Q}_\alpha=(\pi,0)$ (or $(0,\pi)$)  because the 'feedback effect' of the gap opening modifies the quasiparticle energies connected by these electron-hole nesting vectors and in fact destroys or reduces the nesting properties for the quasiparticle bands. Therefore deep in the SDW phase the RKKY function will more resemble the one in Fig.~(\ref{fig5}.d) with the maximum at the $\Gamma$ point. 

\begin{figure}
 \centering
    \subfigure[]
    {
        \includegraphics[width= 0.33\linewidth]{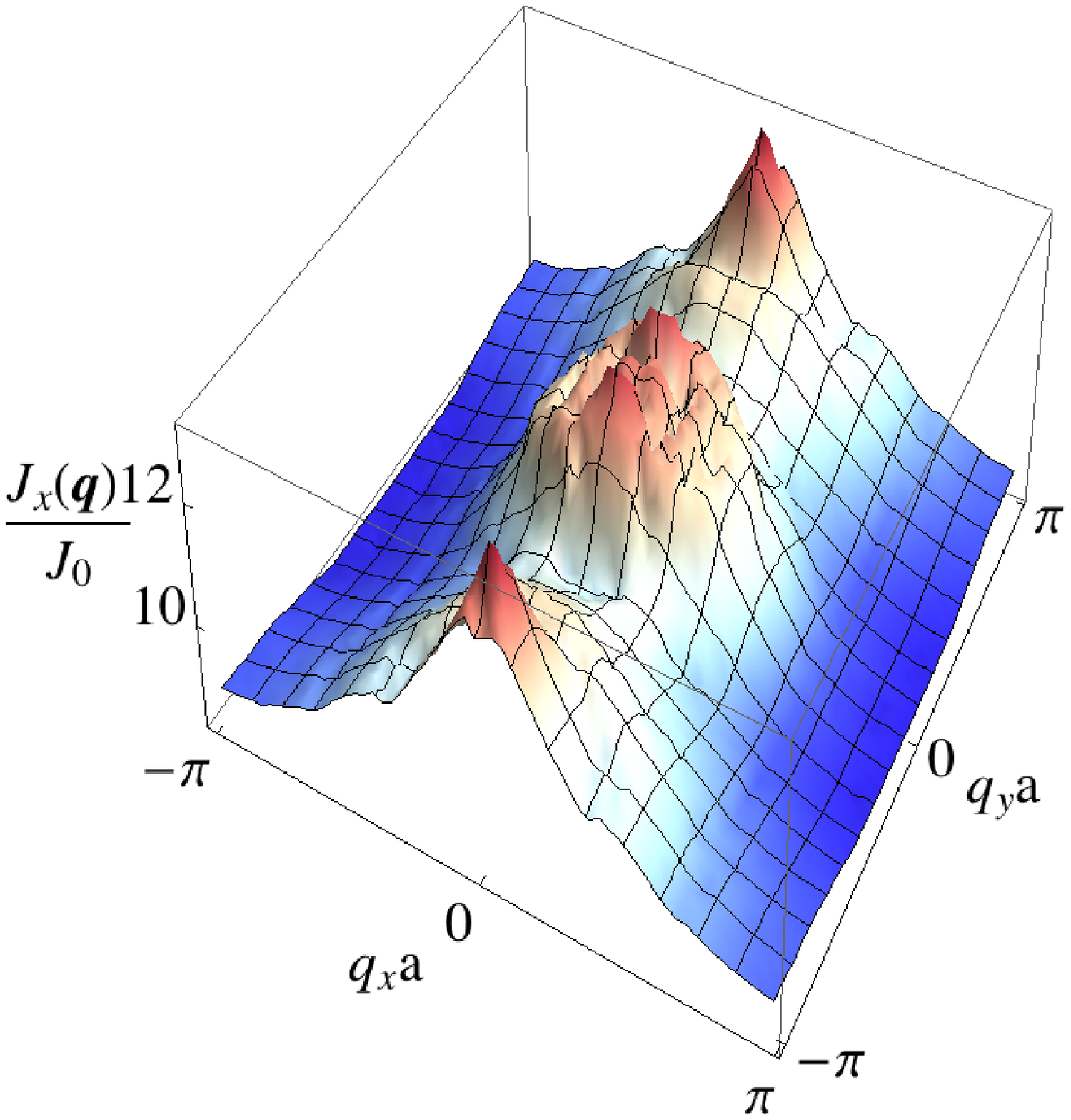}
    }
        \hspace{0.25cm}
    \subfigure[]
    {
        \includegraphics[width= 0.33\linewidth]{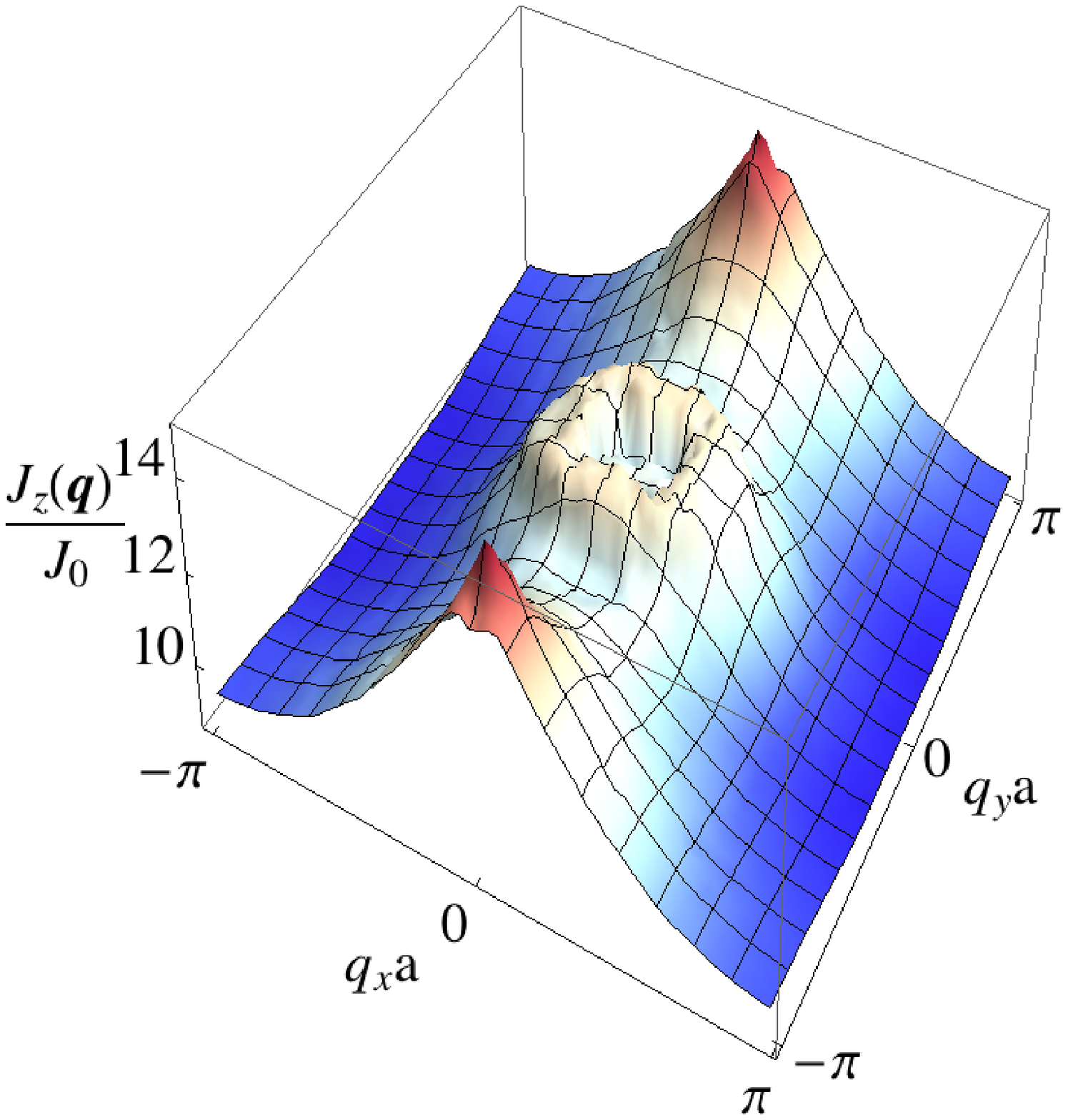}
    }
\caption{(Color online)
 The momentum dependence of the Fourier transform of the  total RKKY interaction in the SDW phase, correspond to a typical tight-binding type band structure  example  for 122 compounds, (see Fig.\ref{fig6}): a) $J_x({\bf q})$, and b) $J_z({\bf q})$.
 \vspace{0.4cm}
 }
\label{fig8}
\end{figure}

This scenario seems indeed to apply to the 4f- based Fe pnictide compound EuFe$_2$As$_2$. There the AF ordering of itinerant 3d moments takes place below $T_{\mbox{SDW}}=190\mbox{K}$ at  ${\bf Q}_\alpha=(\pi,0,0),(0,\pi,0)$ wave vectors \cite{Xiao:09}. It is followed by the AF ordering of localized $Eu^{2+} (S=\frac{7}{2})$ 4f- moments at a much lower temperature $T_N\approx 20\mbox{K}$ and at a {\it different} wave vector ${\bf q}=(0,0,1)$ which means the $Eu^{2+}$ ab-planes with 4f moments are {\it ferromagnetically} ordered in contrast to the columnar AF order of Fe itinerant 3d moments. This agrees with the arguments given above that the RKKY interaction within the SDW phase is dominated by the broad peak near the $\Gamma$ -point as presented in Fig.~(\ref{fig5}.d). 
This can be fully understood by the feedback effect on the \bq-dependence of the RKKY interaction in the presence of the SDW ordering (Eq.(\ref{RKKY_q_SDW})).
The SDW couples  electron (centered at ${\bf Q}_1$) and hole pockets together which leads to reconstructed quasiparticle bands (Sec.\ref{subsect:RKKY_latticeSDW}). Their associated FS has new small pockets around the $\Gamma$-point shown in Fig.(\ref{fig6}.d).
This causes the RKKY contributions of the involved pockets and the maximum to move to the zone center.
The feedback effect of the SDW not only shifts the RKKY maximum to the zone center but also leads to an effective induced RKKY spin space anisotropy as argued in Ref.~\cite{Akbari:2011fk} and experimentally supported in Refs.~\cite{Zapf:11,Dengler:10}. This is also observed in the \bq-dependence of RKKY as presented in Fig.(\ref{fig8}.a) and Fig.(\ref{fig8}.b) for $J_x$ and $J_z$ respectively.
 
\section{Summary}
\label{sect:summary}

We have investigated the RKKY interaction mechanism in 3d multiband Fe pnictide compounds which is relevant for effective coupling
of localized, e.g., 4f moments. We used a simple parabolic 
band model which allows to describe pure electron, hole and composite Fermi surface models. In the former cases a closed
analytical solution for the effective RKKY function was obtained and it is in full agreement with numerical calulation.
In the latter case with both electron and hole sheets only numerical evaluation is possible.
The RKKY interaction is determined by the sum of two intra-band (e-e and h-h) and two inter-band (e-e and e-h) contributions. The former lead to slow spatial oscillations determined by the pocket size, the latter to superposed rapid oscillations determined by
the inter-pocket nesting vectors. Depending on which Fermi surface sheets are present only the former or both type of
oscillations are present in the total RKKY interaction (Fig.~\ref{fig3}).

 We also studied the momentum dependence of the RKKY interaction which would determine the magnetic order in 4f-based
Fe pnictides in a purely 2D picture. We found that in the  parabolic band  model the maximum of the interaction is at the
zone center leading to ferromagnetic order of rare earth planes in these compounds in agreement with observation. A more closer
investigation shows that the RKKY interaction derived from a more realistic tight-binding band model would predict a columnar AF
order in the plane as for the itinerant Fe moments. However, the feedback mechanism of Fe SDW gap opening on the RKKY interaction shifts the ordering vector to the zone center as obtained in the simplified band model.

It is worthwile to discuss to what extent this purely 2D model analysis may be relevant for the ordering in the 4f based iron pnictides like RFe$_2$As$_2$ and RFeAsO. In particular in EuFe$_2$As$_2$   \cite{Xiao:09}  and CeFeAsO the 4f magnetic order has been investigated. The former is most directly related to the present model because this 122 structure has   concerning the local moment properties  planar Eu layers. Furthermore the $Eu^{2+}$ state has a pure spin $S=7/2$ without complications due to crystalline electric field effects and therefore should closely correspond to the effective spin exchange model. In fact this compound exhibits the usual columnar AF SDW order for Fe moments while a much lower T$_N$=20 K the Eu- moments in each tetragonal ab plane order {\it ferromagnetically} as predicted by the present effective exchange model and the feedback effect. The stacking along c which is not described in our context is still antiferromagnetic but may also become ferromagnetic by substitution of As with P \cite{Zapf:11}.
The 1111- structure corresponds less ideally to our model because the Ce atoms do not reside in planar layers. The structure suggests an increased importance of Fe-Ce interlayer coupling which would tend to enforce the columnar AF structure also on the
Ce layers. In fact there are large polarization effects of Ce-moments above the Ce ordering temperature due to this coupling \cite{Maeter:2009}. On the other hand the feedback effect still favors ferromagnetic effective exchange within the Ce layers. The resulting non-collinear Ce-ordering below  $T_N$=4 K seems to be a compromise between these two effects; therefore the moments of ferromagnetic  Ce columns along b are not opposite along a as in the Fe layers but only perpendicular. We conclude that at least in these two examples the tendency to ferromagnetic effective in-plane exchange found in our model seems to be relevant for the real ordering of 4f moments.

%
 \section*{Acknowledgments}

We would like to  thank D. S. Inosov for useful  and stimulating discussions on this work.
We are grateful to the Max Planck Institute for the Physics of Complex Systems (MPI-PKS) for the use of computer facilities.
A.A. would like to thank the Abdus Salam International Centre for Theoretical Physics (ICTP) for hospitality.
 I.E. is supported by the Dresden Platform for
Superconductivity and Magnetism.

\section*{References}
\bibliographystyle{NJP}
\bibliography{RKKY}

\begin{thebibliography}{10}

\bibitem{Kamihara:2008fk}
Y.~Kamihara, T.~Watanabe, M.~Hirano, and H.~Hosono, 2008 \emph{Journal of the
  American Chemical Society}, 130(11) 3296--3297.

\bibitem{Ishida:2009}
K.~Ishida, Y.~Nakai, and H.~Hosono, 2009 \emph{Journal of the Physical Society
  of Japan}, 78(6) 062001.

\bibitem{Stewart:2011}
G.~R. Stewart, 2011 \emph{Rev. Mod. Phys.}, 83 1589--1652.

\bibitem{Lumsden:2010}
M.~D. Lumsden and A.~D. Christianson, 2010 \emph{Journal of Physics: Condensed
  Matter}, 22(20) 203203.

\bibitem{Xiao:09}
Y.~Xiao, Y.~Su, M.~Meven, R.~Mittal, C.~M. Kumar, T.~Chatterji, S.~Price,
  J.~Persson, N.~Kumar, S.~K. Dhar, A.~Thamizhavel, and T.~Brueckel, 2009
  \emph{Phys. Rev B}, 80 174424.

\bibitem{Zhao:2008b}
J.~Zhao, Q.~Huang, C.~de~la Cruz, S.~Li, J.~W. Lynn, Y.~Chen, M.~A. Green,
  G.~F. Chen, G.~Li, Z.~Li, J.~L. Luo, N.~L. Wang, and P.~Dai, 2008 \emph{Nat
  Mater}, 7(12) 953--959.

\bibitem{Maeter:2009}
H.~Maeter, H.~Luetkens, Y.~G. Pashkevich, A.~Kwadrin, R.~Khasanov, A.~Amato,
  A.~A. Gusev, K.~V. Lamonova, D.~A. Chervinskii, R.~Klingeler, C.~Hess,
  G.~Behr, B.~B\"uchner, and H.-H. Klauss, 2009 \emph{Phys. Rev. B}, 80 094524.

\bibitem{Pourovskii:2008}
L.~Pourovskii, V.~Vildosola, S.~Biermann, and A.~Georges, 2008 \emph{EPL
  (Europhysics Letters)}, 84(3) 37006.

\bibitem{Jesche:2009}
A.~Jesche, C.~Krellner, M.~de~Souza, M.~Lang, and C.~Geibel, 2009 \emph{New
  Journal of Physics}, 11(10) 103050.

\bibitem{Kitchen:2006}
D.~Kitchen, A.~Richardella, J.-M. Tang, M.~E. Flatte, and A.~Yazdani, 2006
  \emph{Nature}, 442(7101) 436--439.

\bibitem{Texier:2012}
Y.~Texier, Y.~Laplace, P.~Mendels, J.~T. Park, G.~Friemel, D.~L. Sun, D.~S.
  Inosov, C.~T. Lin, and J.~Bobroff, 2012 \emph{EPL (Europhysics Letters)},
  99(1) 17002.

\bibitem{Alfonsov:2012}
A.~Alfonsov, F.~Mur{\'a}nyi, N.~Leps, R.~Klingeler, A.~Kondrat, C.~Hess,
  S.~Wurmehl, A.~K{\"o}filer, G.~Behr, V.~Kataev, and B.~B{\"u}chner, 2012
  \emph{Journal of Experimental and Theoretical Physics}, 114(4) 662.

\bibitem{Jeevan:2008}
H.~S. Jeevan, Z.~Hossain, D.~Kasinathan, H.~Rosner, C.~Geibel, and
  P.~Gegenwart, 2008 \emph{Phys. Rev. B}, 78 052502.

\bibitem{Ren:2009}
Z.~Ren, Q.~Tao, S.~Jiang, C.~Feng, C.~Wang, J.~Dai, G.~Cao, and Z.~Xu, 2009
  \emph{Phys. Rev. Lett.}, 102 137002.

\bibitem{Jiang:2009}
S.~Jiang, Y.~Luo, Z.~Ren, Z.~Zhu, C.~Wang, X.~Xu, Q.~Tao, G.~Cao, and Z.~Xu,
  2009 \emph{New Journal of Physics}, 11(2) 025007.

\bibitem{Jeevan:2011}
H.~S. Jeevan, D.~Kasinathan, H.~Rosner, and P.~Gegenwart, 2011 \emph{Phys. Rev.
  B}, 83 054511.

\bibitem{Tokiwa:2012}
Y.~{Tokiwa}, S.-H. {H{\"u}bner}, O.~{Beck}, H.~S. {Jeevan}, and P.~{Gegenwart},
  2012 \emph{ArXiv e-prints}.

\bibitem{Dengler:10}
E.~Dengler, J.~Deisenhofer, H.-A. Krug~von Nidda, S.~Khim, J.~S. Kim, K.~H.
  Kim, F.~Casper, C.~Felser, and A.~Loidl, 2010 \emph{Phys. Rev. B}, 81 024406.

\bibitem{Zapf:11}
S.~Zapf, D.~Wu, L.~Bogani, H.~S. Jeevan, P.~Gegenwart, and M.~Dressel, 2011
  \emph{Phys. Rev B}, 84 140503(R).

\bibitem{Tian:2010}
W.~Tian, W.~Ratcliff, M.~G. Kim, J.-Q. Yan, P.~A. Kienzle, Q.~Huang, B.~Jensen,
  K.~W. Dennis, R.~W. McCallum, T.~A. Lograsso, R.~J. McQueeney, A.~I. Goldman,
  J.~W. Lynn, and A.~Kreyssig, 2010 \emph{Phys. Rev. B}, 82 060514.

\bibitem{Ruderman:1954vn}
M.~A. Ruderman and C.~Kittel, 1954 \emph{Phys. Rev.}, 96 99--102.

\bibitem{Kasuya:1956fr}
T.~Kasuya, 1956 \emph{Progress of Theoretical Physics}, 16(1) 45--57.

\bibitem{Yosida:1957zr}
K.~Yosida, 1957 \emph{Phys. Rev.}, 106 893--898.

\bibitem{Aristov:1997qy}
D.~N. Aristov and S.~V. Maleyev, 1997 \emph{Phys. Rev. B}, 56 8841--8848.

\bibitem{Yaresko:2009}
A.~N. Yaresko, G.-Q. Liu, V.~N. Antonov, and O.~K. Andersen, 2009 \emph{Phys.
  Rev B}, 79 144421.

\bibitem{Coldea:2008}
A.~I. Coldea, J.~D. Fletcher, A.~Carrington, J.~G. Analytis, A.~F. Bangura,
  J.-H. Chu, A.~S. Erickson, I.~R. Fisher, N.~E. Hussey, and R.~D. McDonald,
  2008 \emph{Phys. Rev. Lett.}, 101 216402.

\bibitem{Singh:2012}
D.~J. Singh, 2012 \emph{Science and Technology of Advanced Materials}, 13(5)
  054304.

\bibitem{Sadovskii:2012}
M.~Sadovskii, E.~Kuchinskii, and I.~Nekrasov, 2012 \emph{Journal of Magnetism
  and Magnetic Materials}, 324(21) 3481 -- 3486.

\bibitem{Lebegue:2007}
S.~Leb\`egue, 2007 \emph{Phys. Rev. B}, 75 035110.

\bibitem{Singh:2008}
D.~J. Singh and M.-H. Du, 2008 \emph{Phys. Rev. Lett.}, 100 237003.

\bibitem{Boeri:2008}
L.~Boeri, O.~V. Dolgov, and A.~A. Golubov, 2008 \emph{Phys. Rev. Lett.}, 101
  026403.

\bibitem{Mazin:2008}
I.~I. Mazin, D.~J. Singh, M.~D. Johannes, and M.~H. Du, 2008 \emph{Phys. Rev.
  Lett.}, 101 057003.

\bibitem{Liu:2008}
C.~Liu, G.~D. Samolyuk, Y.~Lee, N.~Ni, T.~Kondo, A.~F. Santander-Syro, S.~L.
  Bud'ko, J.~L. McChesney, E.~Rotenberg, T.~Valla, A.~V. Fedorov, P.~C.
  Canfield, B.~N. Harmon, and A.~Kaminski, 2008 \emph{Phys. Rev. Lett.}, 101
  177005.

\bibitem{Evtushinsky:2009}
D.~V. Evtushinsky, D.~S. Inosov, V.~B. Zabolotnyy, A.~Koitzsch, M.~Knupfer,
  B.~B\"uchner, M.~S. Viazovska, G.~L. Sun, V.~Hinkov, A.~V. Boris, C.~T. Lin,
  B.~Keimer, A.~Varykhalov, A.~A. Kordyuk, and S.~V. Borisenko, 2009
  \emph{Phys. Rev. B}, 79 054517.

\bibitem{Xu:2008kx}
G.~Xu, H.~Zhang, X.~Dai, and Z.~Fang, 2008 \emph{EPL (Europhysics Letters)},
  84(6) 67015.

\bibitem{Sato:2009yq}
T.~Sato, K.~Nakayama, Y.~Sekiba, P.~Richard, Y.-M. Xu, S.~Souma, T.~Takahashi,
  G.~F. Chen, J.~L. Luo, N.~L. Wang, and H.~Ding, 2009 \emph{Phys. Rev. Lett.},
  103 047002.

\bibitem{Guo:2010uq}
J.~Guo, S.~Jin, G.~Wang, S.~Wang, K.~Zhu, T.~Zhou, M.~He, and X.~Chen, 2010
  \emph{Phys. Rev. B}, 82 180520.

\bibitem{Qian:2011qy}
T.~Qian, X.-P. Wang, W.-C. Jin, P.~Zhang, P.~Richard, G.~Xu, X.~Dai, Z.~Fang,
  J.-G. Guo, X.-L. Chen, and H.~Ding, 2011 \emph{Phys. Rev. Lett.}, 106 187001.

\bibitem{Szalowski:2008}
K.~Szalowski and T.~Balcerzak, 2008 \emph{Phys. Rev. B}, 78 024419.

\bibitem{Smirnov:2009}
S.~Smirnov, 2009 \emph{Phys. Rev. B}, 79 134403.

\bibitem{Smirnov:2010}
S.~Smirnov, 2010 \emph{Phys. Rev. B}, 81 214425.

\bibitem{Akbari:2010lr}
A.~Akbari, I.~Eremin, and P.~Thalmeier, 2010 \emph{Phys. Rev. B}, 81 014524.

\bibitem{Akbari:2011fk}
A.~Akbari, I.~Eremin, and P.~Thalmeier, 2011 \emph{Phys. Rev. B}, 84 134513.

\bibitem{Aristov:1997ly}
D.~N. Aristov, 1997 \emph{Phys. Rev. B}, 55 8064--8066.

\bibitem{Schwabe:1996uq}
N.~F. Schwabe, R.~J. Elliott, and N.~S. Wingreen, 1996 \emph{Phys. Rev. B}, 54
  12953--12968.

\bibitem{Abramowitz:1984}
M.~Abramowitz and I.~Stegun, editors, 1984 \emph{Handbook of Mathematical
  Functions}.
\newblock Verlag Harri Deutsch, Thun-Frankfurt am Main.

\bibitem{Akbari:2010}
A.~Akbari, J.~Knolle, I.~Eremin, and R.~Moessner, 2010 \emph{Phys. Rev. B}, 82
  224506.

\bibitem{Eremin:2010lr}
I.~Eremin and A.~V. Chubukov, 2010 \emph{Phys. Rev. B}, 81 024511.

\end{thebibliography}

\end{document}